\documentclass[acmlarge]{acmart}
\usepackage{booktabs}   %% For formal tables:
\usepackage{subcaption} %% For complex figures with subfigures/subcaptions
\usepackage{amsmath}
\usepackage{amssymb}
\usepackage[plain,Figure]{algorithm}
\usepackage[noend]{algpseudocode}
\usepackage{graphicx}
\usepackage{stmaryrd}
\usepackage{mathtools}
\usepackage{xspace}
\usepackage{color}
\usepackage{comment}
\usepackage{multirow}
\usepackage{amsthm}
\usepackage{array}
\usepackage{booktabs}
\usepackage{pifont}
\usepackage{bbold}

\usepackage{wrapfig}
\usepackage{hyperref}

\makeatletter\if@ACM@journal\makeatother
\acmJournal{PACMPL}
\acmVolume{1}
\acmNumber{1}
\acmArticle{1}
\acmYear{2017}
\acmMonth{1}
\acmDOI{10.1145/nnnnnnn.nnnnnnn}
\startPage{1}
\else\makeatother
\acmConference[PL'17]{ACM SIGPLAN Conference on Programming Languages}{January 01--03, 2017}{New York, NY, USA}
\acmYear{2017}
\acmISBN{978-x-xxxx-xxxx-x/YY/MM}
\acmDOI{10.1145/nnnnnnn.nnnnnnn}
\startPage{1}
\fi

\setcopyright{none}             %% For review submission
\begin{document}

%% Title information
%%\title[Short Title]{Full Title}         %% [Short Title] is optional;
%\title{Automating Data Completion Tasks from Examples}         %% [Short Title] is optional;

\title{Synthesis of Data Completion Scripts using Finite Tree Automata}
                                        %% when present, will be used in
                                        %% header instead of Full Title.
%%\titlenote{with title note}             %% \titlenote is optional;
                                        %% can be repeated if necessary;
                                        %% contents suppressed with 'anonymous'
%%\subtitle{Subtitle}                     %% \subtitle is optional
%%\subtitlenote{with subtitle note}       %% \subtitlenote is optional;
                                        %% can be repeated if necessary;
                                        %% contents suppressed with 'anonymous'

%% Author information
%% Contents and number of authors suppressed with 'anonymous'.
%% Each author should be introduced by \author, followed by
%% \authornote (optional), \orcid (optional), \affiliation, and
%% \email.
%% An author may have multiple affiliations and/or emails; repeat the
%% appropriate command.
%% Many elements are not rendered, but should be provided for metadata
%% extraction tools.

%% Author with single affiliation.
\author{Xinyu Wang}
%%\authornote{with author1 note}          %% \authornote is optional;
                                        %% can be repeated if necessary
%%\orcid{nnnn-nnnn-nnnn-nnnn}             %% \orcid is optional
\affiliation{
  %%\position{Position1}
  %%\department{Department1}              %% \department is recommended
  \institution{UT Austin}            %% \institution is required
  %%\streetaddress{Street1 Address1}
  %%\city{City1}
  %%\state{State1}
  %%\postcode{Post-Code1}
  %%\country{Country1}
}
\email{xwang@cs.utexas.edu}          %% \email is recommended

%% Author with two affiliations and emails.
\author{Isil Dillig}
%%\authornote{with author2 note}          %% \authornote is optional;
                                        %% can be repeated if necessary
%%\orcid{nnnn-nnnn-nnnn-nnnn}             %% \orcid is optional
\affiliation{
  %%\position{Position2a}
  %%\department{Department2a}             %% \department is recommended
  \institution{UT Austin}           %% \institution is required
  %%\streetaddress{Street2a Address2a}
  %%\city{City2a}
  %%\state{State2a}
  %%\postcode{Post-Code2a}
  %%\country{Country2a}
}
\email{isil@cs.utexas.edu}         %% \email is recommended

%% Author with two affiliations and emails.
\author{Rishabh Singh}
%%\authornote{with author2 note}          %% \authornote is optional;
                                        %% can be repeated if necessary
%%\orcid{nnnn-nnnn-nnnn-nnnn}             %% \orcid is optional
\affiliation{
  %%\position{Position2a}
  %%\department{Department2a}             %% \department is recommended
  \institution{Microsoft Research}           %% \institution is required
  %%\streetaddress{Street2a Address2a}
  %%\city{City2a}
  %%\state{State2a}
  %%\postcode{Post-Code2a}
  %%\country{Country2a}
}
\email{risin@microsoft.com}         %% \email is recommended

%% Paper note
%% The \thanks command may be used to create a "paper note" ---
%% similar to a title note or an author note, but not explicitly
%% associated with a particular element.  It will appear immediately
%% above the permission/copyright statement.
%\thanks{with paper note}                %% \thanks is optional
                                        %% can be repeated if necesary
                                        %% contents suppressed with 'anonymous'

%% Abstract
%% Note: \begin{abstract}...\end{abstract} environment must come
%% before \maketitle command

%%%%%%% Macros for language %%%%%%%
\newcommand{\bottom}{\perp}
\newcommand{\assign}{:=}
\newcommand{\ssemantics}[1]{\semantics{#1}~\state}
\newcommand{\langif}{\text{if}}
\newcommand{\bindto}{\shortleftarrow}
\newcommand{\dsl}{\text{DSL}}
\newcommand{\llangle}{\scalebox{1.4}[0.4]{\raisebox{.4ex}{$\langle$}}}
\newcommand{\rrangle}{\scalebox{1.4}[0.4]{\raisebox{.4ex}{$\rangle$}}}
\newcommand{\inputtable}{\t{T}}
\newcommand{\fst}{\t{row}}
\newcommand{\snd}{\t{col}}
\newcommand{\updir}{\t{u}}
\newcommand{\downdir}{\t{d}}
\newcommand{\leftdir}{\t{l}}
\newcommand{\rightdir}{\t{r}}
\newcommand{\tab}{~~~~}
\newcommand{\constcell}{c}
\newcommand{\seqprog}{\pi}
\newcommand{\unionprog}{\rho}
\newcommand{\simpleprog}{\rho}
\newcommand{\cellprog}{\tau}
\newcommand{\inputcell}{x}
\newcommand{\itercell}{z}
\newcommand{\startcell}{y}
\newcommand{\seqconstruct}{\t{Seq}}
\newcommand{\unionconstruct}{\t{List}}
\newcommand{\filterconstruct}{\t{Filter}}
\newcommand{\getcell}{\t{GetCell}}
\newcommand{\true}{\t{True}}
\newcommand{\tvalue}{\t{Val}}
\newcommand{\valeq}[2]{\tvalue(#1) = \tvalue(#2)}
\newcommand{\valeqconst}[2]{\tvalue(#1) = #2}
\newcommand{\valneqconst}[2]{\tvalue(#1) \neq #2}
\newcommand{\str}{s}
\newcommand{\x}{c}
\newcommand{\idmapper}{\lambda \x. \x}
\newcommand{\rowmapper}[1]{\lambda \x. (#1, \snd(\x))}
\newcommand{\colmapper}[1]{\lambda \x. (\fst(\x), #1)}
\newcommand{\assert}{\t{assert}}
\newcommand{\setand}{\cap}
\newcommand{\setor}{\cup}
\newcommand{\bigsetor}{\bigcup}
\newcommand{\conj}{\wedge}
\newcommand{\disj}{\vee}
\newcommand{\mapper}{\chi}
\newcommand{\semantics}[1]{\llbracket{#1}\rrbracket}
\newcommand{\starunion}{\uplus}
\newcommand{\lift}[1]{\overline{#1}}

%%%%%%% Macros for Algorithms %%%%%%%
\newcommand{\learnseq}{LearnExtractor}
\newcommand{\learnsimple}{LearnSimpProg}
\newcommand{\continue}{\textbf{continue}}
\newcommand{\score}{\theta}
\newcommand{\dom}{dom}
\newcommand{\project}{\downarrow}
\newcommand{\buildprogset}{FTA}
\newcommand{\intersect}{\setand}
\newcommand{\rank}{\textsc{Rank}}
\newcommand{\examples}{\mathcal{E}}
\newcommand{\learn}{Learn}
\newcommand{\failure}{\text{null}}
\newcommand{\dir}{\t{dir}}
\newcommand{\pred}{\phi}
\newcommand{\irule}[2]{\mkern-2mu\displaystyle\frac{#1}{\vphantom{,}#2}\mkern-2mu}
\newcommand{\extresult}{\mathcal{C}}
\newcommand{\sketch}{\mathcal{S}}

%%%%%%% Macros for Paper %%%%%%%%%%
\renewcommand{\t}[1]{\textsf{#1}}
\newcommand{\langtext}[1]{\text{#1}}
\newcommand{\tool}{\textsc{DACE}\xspace}
\renewcommand{\dots}{\cdot\cdot}
\renewcommand{\ldots}{\dots}
\newcommand{\todo}[1]{{\color{red}{#1}}}

\renewcommand{\algref}[1]{Algorithm~\ref{alg:#1}}
\newcommand{\alglabel}[1]{\label{alg:#1}}
\newcommand{\figref}[1]{Fig.~\ref{fig:#1}}
\newcommand{\figlabel}[1]{\label{fig:#1}}
\newcommand{\exref}[1]{Example~\ref{exp:#1}}
\newcommand{\exlabel}[1]{\label{exp:#1}}
\newcommand{\defref}[1]{Definition~\ref{def:#1}}
\newcommand{\deflabel}[1]{\label{def:#1}}
\newcommand{\tabref}[1]{Table~\ref{def:#1}}
\newcommand{\tablabel}[1]{\label{def:#1}}
\newcommand{\lemref}[1]{Lemma~\ref{def:#1}}
\newcommand{\lemlabel}[1]{\label{def:#1}}

\newcommand{\andsym}{\emph{and}}
\newcommand{\orsym}{\emph{or}}

\newcommand{\david}{\text{David}}

\newcommand{\missingval}{\textnormal{\t{?}}}
\newcommand{\sketchhole}{\textnormal{\texttt{?}}}

\newcommand{\specialcell}[2][c]{%
\begin{tabular}[#1]{@{}c@{}}#2\end{tabular}}
\newcommand{\ex}{\mathcal{E}}
\definecolor{gray}{rgb}{0.5, 0.5, 0.5}
\newcommand{\linenumber}[1]{{\color{gray} \footnotesize #1}}
\newcommand\rott[2]{\rotatebox[origin=c]{#1}{#2}}
\newcommand\rot[1]{\rotatebox[origin=c]{90}{#1}}

%%%%%%% Macros for Representation %%%%%%%%%%
\newcommand{\hypergraph}{\mathcal{H}}
\newcommand{\hyperautomaton}{\mathcal{A}}
\newcommand{\progset}{\mathcal{A}}
\newcommand{\cartesianproduct}{\square}
\newcommand{\squareproduct}{\blacksquare}
\newcommand{\states}{Q}
\newcommand{\finalstate}{q^\star}
\newcommand{\finalstates}{Q_f}
\newcommand{\alphabet}{\mathcal{F}}
\newcommand{\transitionrules}{\Delta}
\newcommand{\rewriteto}{\rightarrow}
\newcommand{\rewritetotrans}{\underset{\mathcal{A}}{\stackrel{\mathclap{\normalfont\mbox{*}}}{\longrightarrow}}}
\renewcommand{\state}{q}
\newcommand{\mappers}{\t{Mappers}}
\newcommand{\preds}{\t{Preds}}
\newcommand{\automaton}{\mathcal{A}}
\newcommand{\lang}{\mathcal{L}}
\newcommand{\cells}{\t{Cells}}
\newcommand{\predicates}{\t{Preds}}
%\newcommand{\mmappers}{\t{Mappers}}

%%%%%%% Macros for Specification %%%%%%%%%%
\newcommand{\holemapsto}{\hookrightarrow}

%%%%%%% Macros for Evaluation %%%%%%%%%%
\newcommand{\timeout}{\text{TO}}
\newcommand{\sketchTool}{\textsc{SKETCH}\xspace}

\newcommand{\andedge}{\texttt{AND}\xspace}
\newcommand{\oredge}{\texttt{OR}\xspace}

\newcommand{\rishabh}[1]{{\color{blue}{#1}}}
\newcommand{\xinyu}[1]{{\color{red}{#1}}}
\newcommand{\isil}[1]{{\color{green}{#1}}}

\newcommand{\prose}{\textsc{PROSE}\xspace}
\newcommand{\complexity}[1]{\mathcal{O}{(#1)}}

\begin{abstract}
In application domains that store data in a tabular format, a common task is to fill the values of some cells using values stored in other cells. For instance, such data completion tasks arise in the context of \emph{missing value imputation} in data science and \emph{derived data} computation in spreadsheets and relational databases. Unfortunately, end-users and data scientists typically struggle with  many data completion tasks that require non-trivial programming expertise. 
This paper presents a synthesis technique for automating data completion tasks using \emph{programming-by-example (PBE)} and a very lightweight sketching approach. Given a \emph{formula sketch} (e.g., {\tt AVG}($\sketchhole_1$, $\sketchhole_2$)) and a few input-output examples for each hole, our technique synthesizes a  program to automate the desired data completion task. Towards this goal, we propose a  domain-specific language (DSL) that combines spatial and relational reasoning over tabular data and a novel synthesis algorithm that can generate DSL programs that are consistent with the input-output examples. The key technical novelty of our approach is a new version space  learning algorithm that is based on \emph{finite tree automata} (FTA). The use of FTAs in the learning algorithm leads to a more compact representation  that allows more sharing between programs that are consistent with the examples. We have implemented the proposed approach in a tool called \tool and evaluate it on 84 benchmarks taken from online help forums. We also illustrate the advantages of our approach by comparing our technique against two existing synthesizers, namely {\sc PROSE} and {\sc SKETCH}.
\end{abstract}

\begin{comment}
%% 2012 ACM Computing Classification System (CSS) concepts
%% Generate at 'http://dl.acm.org/ccs/ccs.cfm'.
\begin{CCSXML}
<ccs2012>
<concept>
<concept_id>10011007.10011006.10011008</concept_id>
<concept_desc>Software and its engineering~General programming languages</concept_desc>
<concept_significance>500</concept_significance>
</concept>
<concept>
<concept_id>10003456.10003457.10003521.10003525</concept_id>
<concept_desc>Social and professional topics~History of programming languages</concept_desc>
<concept_significance>300</concept_significance>
</concept>
</ccs2012>
\end{CCSXML}

\ccsdesc[500]{Software and its engineering~General programming languages}
\ccsdesc[300]{Social and professional topics~History of programming languages}
%% End of generated code
\end{comment}

%% Keywords
%% comma separated list
%%\keywords{keyword1, keyword2, keyword3}  %% \keywords is optional

%% \maketitle
%% Note: \maketitle command must come after title commands, author
%% commands, abstract environment, Computing Classification System
%% environment and commands, and keywords command.
\maketitle

\section{Introduction}\label{sec:intro}

Many application domains store  data in a {tabular form} arranged using rows and columns. For example, Excel spreadsheets, R dataframes, and relational databases  all view the underlying data as a 2-dimensional table consisting of \emph{cells}. In this context, a common scenario is to fill the values of some cells using values stored in other cells. For instance, consider the following common data completion tasks:

\begin{itemize}
\item {\bf \emph{Data imputation:}} In statistics, \emph{imputation} means replacing missing data with substituted values. Since missing values can hinder data analytics tasks, users often need to fill missing values using other related entries in the table. For instance, data imputation  arises frequently in statistical computing frameworks, such as R and \emph{pandas}.
\item  {\bf \emph{Spreadsheet computation:}} In many applications involving spreadsheets, users need to calculate the value of a cell based on values from other cells. For instance, a common task is to introduce new columns, where each value in the new column is derived from values in existing columns.
\item {\bf \emph{Virtual columns in databases:}} In relational databases, users sometimes create \emph{views} that store the result of some database query. In this context, a common task is to add \emph{virtual columns} whose values are computed using existing entries in the view. 
\end{itemize}

As illustrated by these examples, users often need to complete missing values in tabular data. While some of these data completion tasks are fairly straightforward, many  others require non-trivial programming knowledge that is beyond the expertise of end-users and data scientists.

To illustrate a typical data completion task,  consider the tabular data shown in \figref{fig:intro-ex}. Here, the table stores measurements for different people during a certain time period, where each row represents a person and each column corresponds to a day. As explained in a StackOverflow post~\footnote{http://stackoverflow.com/questions/30952426/substract-last-cell-in-row-from-first-cell-with-number}, a data scientist analyzing this data wants to compute the difference of the measurements between the first and last days for each person and record this information in the \emph{Delta} column. Since the table contains a large number of rows (of which only a small subset is shown in \figref{fig:intro-ex}), manually computing this data is prohibitively cumbersome. Furthermore, since each person's start and end date is different, automating this data completion requires non-trivial programming logic.

\begin{figure}[!t]
\begin{center}
\includegraphics[scale=0.63]{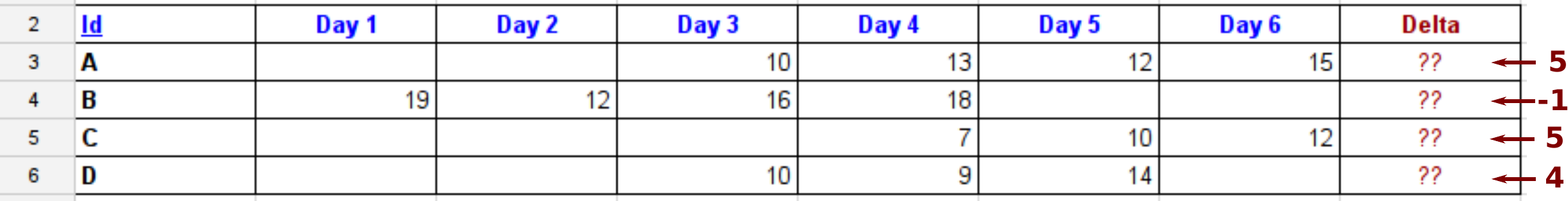}
\end{center}
\vspace{-0.1in}
\caption{A data completion task taken from StackOverflow.}
\figlabel{fig:intro-ex}
\end{figure}

In this paper, we present a novel program synthesis technique for automating data completion tasks in tabular data sources, such as dataframes, spreadsheets, and relational databases. Our synthesis methodology is based on two key insights that we gained by analyzing dozens of posts on online forums: First, it is often {easy} for end-users to specify \emph{which operators} should be used in the data completion task and  provide a specific instantiation of the operands for a few example cells. However,  it is typically {very difficult} for end-users to express the \emph{general operand extraction logic}. For instance, for the example from \figref{fig:intro-ex}, the user knows that the missing value can be computed as $C_1 - C_2$, but he is not sure how to implement the logic for extracting $C_1, C_2$ in the general case.

Based on this observation, our synthesis methodology for  data completion  combines \emph{program sketching} and \emph{programming-by-example (PBE)}. Specifically, given a \emph{ formula sketch} (e.g.,
{\tt
SUM($\sketchhole_1$,AVG($\sketchhole_2$,$\sketchhole_3$))}) as well as a few input-output examples for each hole, our technique automatically synthesizes a program that can be used to fill \emph{all} missing values in the table. For instance, in our running example, the user provides the sketch {\tt MINUS($\sketchhole_1$,$\sketchhole_2$)} as well as the following input-output examples for the two holes:
\vspace{0.1in}

{\small
\begin{center}
\begin{tabular}{|c|c|} 
\hline
$\sketchhole_1$ & $\sketchhole_2$ \\ \hline
(A, Delta) $\mapsto$ (A, Day 6) & (A, Delta) $\mapsto$ (A, Day 3) \\ \hline
(B, Delta) $\mapsto$ (B, Day 4) & (B, Delta) $\mapsto$ (B, Day 1) \\ \hline
\end{tabular}
\end{center}
}
\vspace{0.1in}

\noindent Given these examples, our technique automatically synthesizes a program that can be used to fill all  values in the \emph{Delta} column in \figref{fig:intro-ex}.

One of the key pillars underlying our synthesis algorithm is the design of
%To  automate  such data completion tasks from examples, we propose a novel synthesis algorithm that generates programs over 
a new domain-specific language (DSL) that is expressive enough to capture most data completion tasks we have encountered, yet lightweight enough to facilitate automation. The programs in our DSL take as input a table and a cell with missing value, and return a list of cells that are used for computing the missing value. Our main insight in designing this DSL is to use abstractions that combine spatial reasoning in the tabular structure with relational reasoning. Spatial reasoning allows the DSL programs to follow structured paths in the 2-dimensional table whereas  relational reasoning allows them to constrain those paths with predicates over cell values. 

\begin{comment}
We present a domain specific language (DSL) for computing the hole values in data completion sketches that is expressive enough to capture most tasks we have encountered. The programs in the DSL take as input a table and a cell with missing value, and return a list of cells as output that are in turn used for computing the missing value. The key idea in the DSL is to use abstractions to combine spatial reasoning in the tabular structure with relational reasoning. The spatial reasoning allows the DSL programs to follow structured paths in the 2-dimensional table whereas the relational reasoning allows them to constrain the paths with predicates over cell values. 
\rishabh{do we need to say more about the DSL here?}
\end{comment}

As shown schematically in \figref{fig:overview}, the high-level structure of our synthesis algorithm is similar to prior techniques that combine \emph{partitioning} with \emph{unification}~\cite{flashfill,hades,alur2015synthesis} . Specifically, partitioning is used to classify the input-output examples into a small number of groups, each of which can  be represented using a conditional-free program in the DSL. In contrast, the goal of unification is to find a conditional-free program that is consistent with each example in the group. The key novelty of our synthesis algorithm is a new unification algorithm based on \emph{finite tree automata} (FTA). 

Our unification procedure can be viewed as a new \emph{version space learning} algorithm~\cite{vs} that succinctly represents a large number of programs. Specifically, a \emph{version space} represents all viable hypotheses that are consistent with a given set of examples, and  prior work on programming-by-example have used so-called \emph{version space algebras} (VSA) to combine simpler version spaces into more complex ones~\cite{vsa,flashfill,flashmeta}. Our use of FTAs for version space learning offers several advantages compared to prior VSA-based techniques such as \prose: First, FTAs represent version spaces more succinctly, without explicitly constructing  individual sub-spaces representing sub-expressions. Hence, our approach avoids the need for finite unrolling of recursive expressions in the DSL. Second, our finite-tree automata are constructed in a forward manner using the DSL semantics. In constrast to VSA-based approaches such as \prose that construct VSAs in a backward fashion starting from the outputs, our approach therefore obviates the need to manually define complex inverse semantics for each DSL construct. As we demonstrate experimentally, our version space learning algorithm using FTAs significantly outperforms the VSA-based learning algorithm used in \prose.

%our unification procedure uses \emph{finite tree automata (FTA)} to represent \emph{version spaces} that succinctly represent a large number of programs in the DSL that are consistent with the specification.

\begin{comment}
 The notion of version space was first introduced by Mitchell~\cite{vs} in the context of concept learning, where the key insight was that a set of consistent hypotheses can be represented succinctly using the most-specific and the most-general hypotheses. This idea was later extended for Programming By Demonstration (PBD) by allowing to represent arbitrary partial orders over hypotheses~\cite{vsa}, and VSA has been used quite extensively recently in Programming By Example (PBE) techniques such as FlashFill~\cite{flashfill} and PROSE~\cite{flashmeta}. However, there are two key differences in using FTA compared to the VSA technique in previous approaches such as PROSE. First, FTA allows for constructing more succinct version spaces without the need for explicit construction of each individual sub-spaces (representing the set of sub-expressions), which avoids the need for finite unrolling of recursive expressions in the DSL. Second, the FTA is constructed in a forward manner using the DSL semantics, whereas PROSE constructs VSAs in a backward fashion starting from the outputs, which in turn requires additional manual effort of defining complex inverse semantics for each DSL function.
\end{comment}

\begin{figure}[!t]
\vspace{-0.05in}
\begin{center}
\includegraphics[scale=0.3]{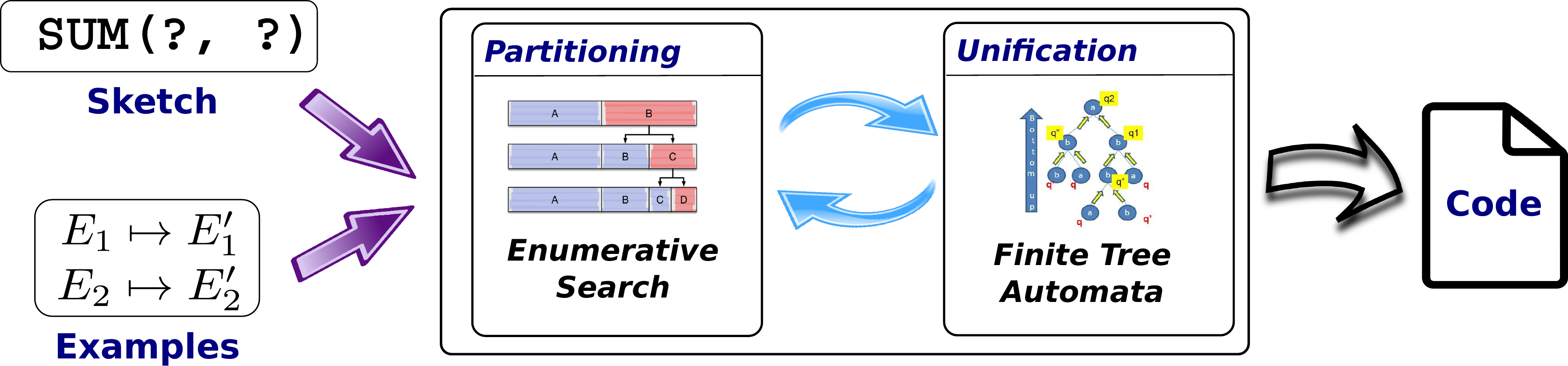}
\end{center}
\vspace{-0.1in}
\caption{High-level overview of our synthesis approach.}\figlabel{fig:overview}
\vspace{-0.2in}
\end{figure}

We have implemented our synthesis algorithm in a tool called \tool~\footnote{\tool stands for \underline{DA}ta \underline{C}ompletion \underline{E}ngine.} and evaluated it on $84$ real-world data completion tasks collected from online help forums. 
Our evaluation shows that \tool can successfully synthesize over $92\%$ of the data completion tasks in an average of $0.7$ seconds. We also empirically compare our approach against \prose and \sketchTool and show that our new synthesis technique outperforms these baseline algorithms  by orders of magnitude in the data completion domain.

To summarize, this paper makes the following key contributions:

\begin{itemize}
\item We propose a new program synthesis methodology that combines \emph{program sketching} and \emph{programming-by-example} techniques  to automate a large class of data completion tasks involving tabular data.
\item We describe a DSL that can concisely express a large class of data completion tasks and  that is amenable to an efficient synthesis algorithm.
\item We propose a new unification algorithm that uses \emph{finite tree automata} to construct version spaces that succinctly represent DSL programs  that are consistent with a set of input-output examples. 
\item We evaluate our approach on real-world data completion tasks involving dataframes, spreadsheets, and relational databases. Our experiments show that the proposed learning algorithm is effective in practice.
\end{itemize}

\section{Motivating Examples}\label{sec:examples}

In this section, we present some representative data completion tasks collected from online help forums. Our main goal here is to demonstrate the diversity of data completion tasks and motivate various design choices for the DSL of cell extraction programs. %presented in Section~\ref{sec:dsl}.

\begin{example}%[LOCF]
A numerical ecologist needs to perform data imputation in R using the \emph{last observation carried forward (LOCF)} method~\footnote{\url{http://stackoverflow.com/questions/38100208/fill-missing-value-based-on-previous-values}}. Specifically, he would like to replace each missing entry with the previous non-missing entry incremented by $1$. For instance, if the original row contains
$[2, 2, \missingval, \missingval, 8, \missingval]$, the new row should be
$[2, 2, 3, 3, 8, 9]$. 
\exlabel{locf}
\end{example}

Here, the desired imputation task can be expressed using the simple formula sketch $\texttt{SUM}(\sketchhole_1,1)$. The synthesis task is to find an expression that retrieves the previous non-missing value for each missing entry.

\begin{example}
An astronomer needs to perform data imputation using Python's \emph{pandas} data analysis library~\footnote{\url{http://stackoverflow.com/questions/16345583/fill-in-missing-pandas-data-with-previous-non-missing-value-grouped-by-key}}. 
Specifically, the astronomer wants to replace each missing value with the previous non-missing value with the \emph{same id}. The desired data imputation task is illustrated in \figref{exfigs}(a).
\exlabel{locfid}
\end{example}

As this example illustrates, some data completion tasks require finding a cell that satisfies a \emph{relational predicate}. In this case, the desired cell with non-missing value must have the same id as the id of the cell with missing value.

\begin{figure*}
\footnotesize
\begin{tabular}{ c | c | c | c }
\begin{minipage}{0.13\linewidth}
\hspace*{-10pt}
\begin{tabular}{ l l l } 
& \linenumber{1} & \linenumber{2} \\ 
\linenumber{1:} & id	& x \\ 
\linenumber{2:} & $1$	& $10$ \\
\linenumber{3:} & $1$	& $\missingval \shortleftarrow 10$ \\ 
\linenumber{4:} & $2$	& $100$ \\
\linenumber{5:} & $2$	& $200$	\\ 
\linenumber{6:} & $1$	& $\missingval \shortleftarrow 10$ \\ 
\linenumber{7:} & $2$	& $\missingval \shortleftarrow 200$ \\ 
\linenumber{8:} & $1$	& $300$ \\ 
& $\dots$ & $\dots$ \\
\end{tabular}
\end{minipage}
&
\begin{minipage}{0.23\linewidth}
\hspace*{-3pt}
\includegraphics[scale=0.33]{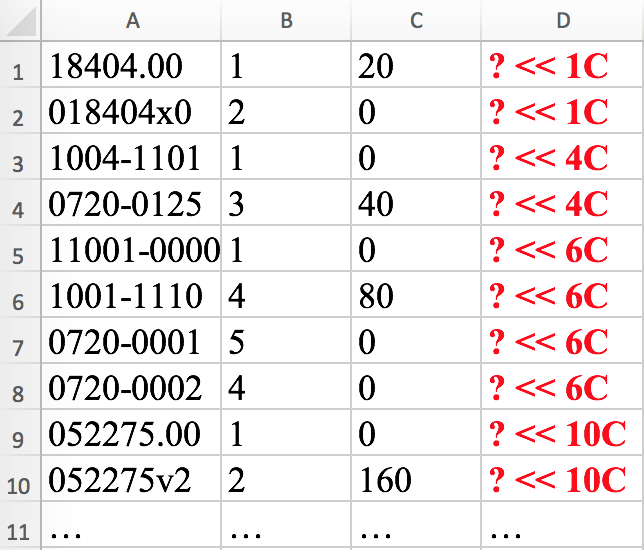}
\end{minipage}
&
\begin{minipage}{0.24\linewidth}
\hspace*{-3pt}
\includegraphics[scale=0.31]{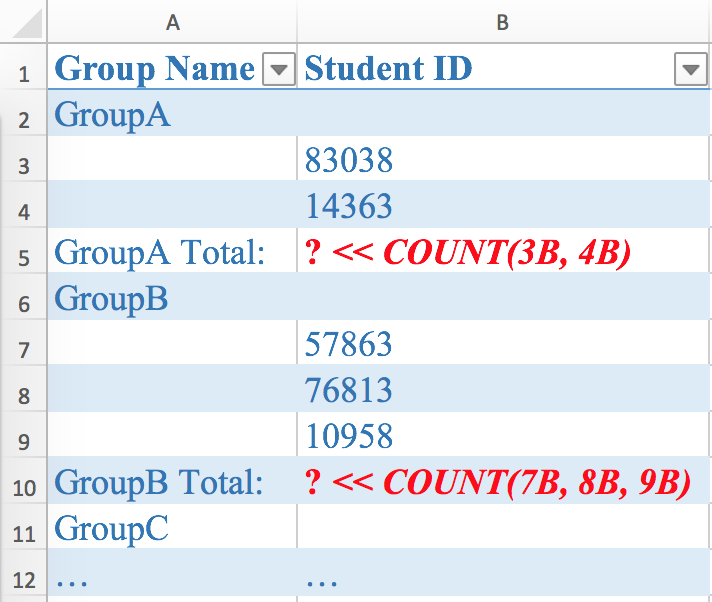}
\end{minipage}
&
\begin{minipage}{0.29\linewidth}
\hspace*{-5pt}
\begin{tabular}{ l l l l l l }
& \linenumber{1} & \linenumber{2} & \linenumber{3} & \linenumber{4} \\ 
\linenumber{1:} & 2016-11-01	& A & $\missingval \shortleftarrow 12$ & $\missingval \shortleftarrow 200$ \\ 
\linenumber{2:} & 2016-11-02	& B & $12$ & $\missingval \shortleftarrow 200$ \\ 
\linenumber{3:} & 2016-11-03	& A & $\missingval \shortleftarrow 12$ & $200$ \\ 
\linenumber{4:} & 2016-11-04	& C & $18$ & $400$ & \\ 
\linenumber{5:} & 2016-11-05	& B & $10$ & $\missingval \shortleftarrow 400$ \\
\linenumber{6:} & 2016-11-06	& B & $\missingval \shortleftarrow 10$ & $800$ \\
\linenumber{7:} & 2016-11-07	& C & $\missingval \shortleftarrow 10$ & $1000$ \\ 
& $\dots$ & $\dots$ & $\dots$ & $\dots$ \\
\end{tabular}
\end{minipage}
\\ 
(a) & (b) & (c) & (d) \\
\end{tabular}
\caption{(a) Replace missing entry by  previous non-missing value with same id. (b) Fill cells in column D by copying values in column C. (c) Calculate total count for each group. (d) Replace missing value by previous non-missing value (if exists), otherwise use next non-missing one.}
\figlabel{exfigs}
\vspace{-5pt}
\end{figure*}

\begin{example}%[Copying data from another column]
A businessman using Excel wants to add a new column to his spreadsheet~\footnote{\url{http://stackoverflow.com/questions/29606616/find-values-between-range-fill-in-next-cells}}. As shown in \figref{exfigs} (b), the entries in the new column D  are obtained from column C. Specifically, the data completion logic is as follows: 
First, find the  previous row that has value  $1$ in column $B$.  Then, go down from that row and find the first non-zero value in column $C$ and use that value to populate the cell in column $D$. \figref{exfigs} (b) shows the desired values for column $D$.
\exlabel{maketurn}
\end{example}

As this example illustrates, the cell extraction logic in some tasks can be quite involved. In this example, we first need to find an \textit{intermediate cell} satisfying a certain property (namely, it must be upwards from the missing cell and have value $1$ in column $B$). Then, once we find this intermediate cell, we need to find the target cell satisfying a different property (namely, it must be downwards from the intermediate cell and store a non-zero value in column $C$). 

This example illustrates two important points: First, the data extraction logic can \textit{combine} both geometric (upward and downward search) and relational properties. Second, the data extraction logic can require ``\textit{making turns}'' --- in this example, we first go up to an intermediate cell and then change direction by going down to find the target cell using a different logic.

\begin{example}%[Sum for each group]
A statistician working on spreadsheet data needs to complete the ``Group Total'' rows from \figref{exfigs} (c) using the count of entries in that group~\footnote{\url{http://stackoverflow.com/questions/13998218/count-values-in-groups/13999520}}. For instance, the \emph{GroupA Total} entry should be filled with $2$.
\end{example}

This example illustrates the need for allowing holes of type \emph{list} in the sketches. Since each missing value is obtained by counting a \emph{variable} number of entries, the user instead provides a sketch $\texttt{COUNT}(\sketchhole_1)$, where $\sketchhole_1$ represents a \emph{list} of cells. Furthermore, this example also illustrates the need for a language construct that can extract a \emph{range} of values satisfying a certain property. For instance, in this example, we need to extract \emph{all} cells between the cell $c$ to be completed and the first empty entry that is upwards from $c$.

\begin{comment}
\begin{figure}[!t]
\footnotesize
\centering
\begin{tabular}{ l l l l l l }
& \linenumber{1} & \linenumber{2} & \linenumber{3} & \linenumber{4} \\ 
\linenumber{1:} & 2016-11-01	& A & $\missing \shortleftarrow 12$ & $\missing \leftarrow 200$ \\ 
\linenumber{2:} & 2016-11-02	& B & $12$ & $\missing \shortleftarrow 200$ \\ 
\linenumber{3:} & 2016-11-03	& A & $\missing \shortleftarrow 12$ & $200$ \\ 
\linenumber{4:} & 2016-11-04	& C & $18$ & $400$ & $\dots $\\ 
\linenumber{5:} & 2016-11-05	& B & $10$ & $\missing \shortleftarrow 400$ \\
\linenumber{6:} & 2016-11-06	& B & $\missing \shortleftarrow 10$ & $800$ \\
\linenumber{7:} & 2016-11-07	& C & $\missing \shortleftarrow 10$ & $1000$ \\ 
& $\dots$ & $\dots$ & $\dots$ & $\dots$ \\
\end{tabular}
\caption{To fill the missing value, first try the previous non-missing value, if it fails, then try the next non-missing value.}
\figlabel{locfseq}
\vspace{-0.1in}
\end{figure}
\end{comment}

\begin{example}%[Conditional LOCF]
An R user wants to perform  imputation on the dataframe shown in \figref{exfigs} (d) using the LOCF method. Specifically, she wants to impute the missing value by substituting it with the first \emph{previous} non-missing value in the same column. However, if no such entry exists, she wants to fill the missing entry by using the first \emph{following} non-missing value in the same column instead.
\exlabel{locfseq} 
\end{example}

This example illustrates the need for allowing a conditional construct in our DSL: Here, we first try some extraction logic, and, if it fails, we then resort to a back-up strategy. Many of the data completion tasks that we have encountered follow this pattern -- i.e., they try different extraction logics depending on whether the previous logic succeeds. Based on this observation, our DSL introduces a restricted form of \emph{switch} statement, where the conditional implicitly checks if the current branch fails.

\begin{comment}
\begin{figure}[!t]
\begin{tabular}{ c | c }
\hspace*{-8pt}
\begin{minipage}{0.47\linewidth}
\includegraphics[scale=0.35]{figs/maketurn.png}
\end{minipage}
&
\begin{minipage}{0.46\linewidth}
\includegraphics[scale=0.32]{figs/filter.png}
\end{minipage}
\\
(a) & (b)
\end{tabular}
\caption{(a) Filling cells in column D by copying values in column C, (b) Calculating total count for each group.}
\figlabel{maketurn}
\vspace{-0.1in}
\end{figure}
\end{comment}

\section{Specifications}

\begin{figure*}[!t]
\small
\centering
\begin{tabular}{| c || c | c| }
\hline
\textbf{Task} & \textbf{Sketch} & \textbf{Input-output examples} \\ \hline
%Figure 1 & $\texttt{MINUS}(\hole_1, \hole_2)$ & \specialcell{$\Big{\{} \hole_1 \holemapsto \{ (A, Delta) \mapsto (A, Day6), (B, Delta) \mapsto (B, Day4) \}$ \\ $~~~~~~\hole_2 \holemapsto \{ (A, Delta) \mapsto (A, Day3), (B, Delta) \mapsto (B, Day1) \}  \Big{\}}$} \\ \hline
Example 2.1 & $\texttt{SUM}(\sketchhole_1, 1)$ & $\big{\{} \sketchhole_1 \holemapsto \{ (1,3) \mapsto [ (1,2) ], (1,4) \mapsto [ (1,2) ], (1,6) \mapsto [ (1,5) ] \} \big{\}}$ \\ \hline
Example 2.2 & $\sketchhole_1$ & $\big{\{} \sketchhole_1 \holemapsto \{ (3,2) \mapsto [ (2,2) ], (6,2) \mapsto [ (2,2) ], (7,2) \mapsto [ (5,2) ] \} \big{\}}$ \\ \hline 
Example 2.3 & $\sketchhole_1$ & $\big{\{} \sketchhole_1 \holemapsto \{ (1,D) \mapsto [ (1,C) ], (3,D) \mapsto [ (4,C) ], (8,D) \mapsto [ (6,C) ] \} \big{\}}$ \\ \hline 
%Example 4 & $\texttt{AVG}(\hole_1, \hole_2)$ & 
%\specialcell{$\Big{\{} 
%\hole_1 \holemapsto \{ (3,2) \mapsto [ (2,2) ], (4,2) \mapsto [ (2,2) ], (7,2) \mapsto [ (6,2) ]$ \\ 
%$~~~~~~~~\hole_2 \holemapsto \{ (3,2) \mapsto [ (5,2) ], (4,2) \mapsto [ (5,2) ], (7,2) \mapsto [ ((8,2) ]\}  \Big{\}}$} \\ \hline
%\begin{array}{c}
%$\Big{\{} \hole_1 \mapsto \{ (3,2) \mapsto [ (2,2) ], (4,3) \mapsto [ (2,2) ], (7,2) \mapsto [ (6,2) ], \\
%\hole_2 \mapsto \{ (3,2) \mapsto [ (5,2) ], (4,3) \mapsto [ (5,2) ], (7,2) \mapsto [ ((8,2) ]\} \Big{\}}$ 
%\end{array} \\ \hline 
Example 2.4 & $\texttt{COUNT}(\sketchhole_1)$ & $\big{\{} \sketchhole_1 \holemapsto \{ (5,B) \mapsto [ (3,B), (4,B) ], (10,B) \mapsto [ (7,B), (8,B), (9,B) ] \} \big{\}}$ \\ \hline 
Example 2.5 & $\sketchhole_1$ & $\big{\{} \sketchhole_1 \holemapsto \{ (3,3) \mapsto [ (2,3) ], (7,3) \mapsto [ (5,3) ], (1,4) \mapsto [ (3,4) ] \} \big{\}}$ \\ \hline 
\end{tabular}
\caption{Specifications for examples from Section~\ref{sec:examples}. }
\figlabel{examplespecs}
\end{figure*}

A specification in our synthesis methodology is a pair $(\sketch, \ex)$, where $\sketch$ is a formula sketch and $\ex$ is a set of input-output examples. Specifically, formula sketches are defined by the following grammar:
\[
\begin{array}{lll}
\langtext{Sketch} ~ \sketch & \assign & t ~ | ~ F(\sketch_1, \dots, \sketch_n), \ \ \ F \in \Lambda \\ 
\langtext{Term} ~ t & \assign & const ~ | ~ \sketchhole_{id} \\ 
\end{array}
\]
Here, $\Lambda$ denotes a family of pre-defined functions, such as {\tt AVG}, {\tt SUM}, {\tt MAX}, etc. Holes in the sketch represent unknown \emph{cell extraction programs} to be synthesized. Observe that formula sketches can contain multiple functions. For instance, ${\tt SUM}({\tt MAX}(\sketchhole_1, \sketchhole_2), 1)$ is a valid sketch and indicates that a missing value in the table should be filled by adding one to the maximum of two unknown cells. %\todo{this does not have to be two cells, because we allow list types?}.

In many cases, the data completion task involves copying values from an existing cell. In this case, the user can express her intent using the identity sketch ${\tt ID}(\sketchhole_1)$. Since this sketch is quite common, we abbreviate it using the notation $\sketchhole_1$. %In other words, the sketch $\hole$ indicates that missing entries should be filled using existing values in the table.

In addition to the sketch, users of \tool are also expected to provide one or more input-output examples $\ex$ for each hole. Specifically, examples $\ex$ map each hole $\sketchhole_{id}$ in the sketch to a set of pairs of the form $i \mapsto [o_1, \ldots, o_n]$, where $i$ is an input cell and $[o_1, \ldots, o_n]$ is the desired list of output cells. Hence, examples in \tool have the following shape:
\[
\langtext{Example} ~ \ex \assign \Big{\{} \sketchhole_{id} \holemapsto \{ i \mapsto [ o_1, \dots, o_n ] \} \Big{\}} \\ 
\]

Each cell in the table is represented as a pair $(x, y)$, where $x$ and $y$ denote the row and column of the cell respectively. \figref{examplespecs} provides the complete specifications for the examples described in Section~\ref{sec:examples}.

Given a specification $(\sketch, \ex)$, the key learning task is to synthesize a program $P_{id}$ for each hole $\sketchhole_{id}$  such that $P_{id}$ satisfies all examples $\ex[\sketchhole_{id}]$. For a list of programs $\mathcal{P} = [P_1, \ldots, P_n]$, we write $\sketch[\mathcal{P}]$ to denote the resulting program that is obtained by filling hole $\sketchhole_{id}$ in $\sketch$ with program $P_{id}$. Once \tool learns a cell extraction program $P_{id}$ for each hole, it computes missing values in the table $\inputtable$ using $\sketch[\mathcal{P}](\inputtable, c)$ where $c$ denotes a cell to be completed in table $\inputtable$. In the rest of the paper, we assume that missing values in the table are identified using the special symbol $\missingval$. For instance, the analog of $\missingval$ is the symbol {\tt NA} in R and blank cell in Excel.

\section{Domain-Specific Language}
\label{sec:dsl}

In this section, we present our domain-specific language (DSL) for cell extraction programs. The syntax of the DSL is shown in \figref{langsyntax}, and its denotational semantics is presented in \figref{langsemantics}. We now review the key constructs in the DSL together with their semantics.

A cell extraction program $\seqprog$ takes as input a table $\inputtable$ and a cell $x$, and returns  a list of cells $[c_1, 
\dots, c_n]$ or the special value $\bot$. Here, $\bot$ can be thought of as an ``exception" and indicates that $\pi$ fails to extract any cells on its input cell $x$. A cell extraction program $\pi$ is either a \emph{simple program} $\rho$ without branches or a conditional of the form  $\seqconstruct(\rho, \pi)$. As shown in \figref{langsemantics}, the semantics of $\seqconstruct(\rho, \pi)$ is that the second argument $\pi$ is only evaluated if $\rho$ fails (i.e., returns~$\bot$).

\begin{figure}[!t]
\vspace*{-8pt}
\[
\small
\begin{array}{lll}
\langtext{Extractor} ~ \seqprog & \assign & \lambda \inputtable.\lambda x. \simpleprog ~~~ | ~~~ \lambda \inputtable.\lambda x. \seqconstruct(\simpleprog, \seqprog) \\ 
\langtext{Simple prog.} ~ \simpleprog & \assign & \unionconstruct(\cellprog_1, \dots, \cellprog_n) ~~~ | ~~~ \filterconstruct(\cellprog_1, \cellprog_2, \cellprog_3, \lambda \startcell. \lambda \itercell. \pred) \\ 
\langtext{Cell prog.} ~ \cellprog & \assign & \inputcell ~~ | ~~ \getcell(\cellprog, \dir, k, \lambda \startcell. \lambda \itercell. \pred) \\ 
\langtext{Predicate} ~ \pred & \assign & \true ~~ | ~~ \valeqconst{\mapper(\itercell)}{s} ~~ | ~~ \valneqconst{\mapper(\itercell)}{s} ~~ | ~~ \valeq{\mapper(\startcell)}{\mapper(\itercell)} ~~ | ~~ \pred_1 \conj \pred_2 \\ 
\langtext{Cell mapper} ~ \mapper & \assign & \idmapper ~~ | ~~ \rowmapper{k} ~~ | ~~ \colmapper{k} \\ 
\langtext{Direction} ~ \dir & \assign & \updir ~ | ~ \downdir ~ | ~ \leftdir ~ | ~ \rightdir \\ 
\end{array}
\]
\vspace{-0.1in}
\caption{Syntax of the $\dsl$. Here,  $\inputtable$ is a table, $\inputcell$ is a cell, and $\startcell$ and $\itercell$ are bound to cells in table $\inputtable$. Also, $k$ denotes an integer constant, and $s$ denotes a string constant. $\tvalue(\cdot)$ returns the value in a given cell.}
\figlabel{langsyntax}
\end{figure}

Let us now consider the syntax and semantics of simple programs $\rho$. A simple program is either a list of individual cell extraction programs ($\unionconstruct(\cellprog_1, \dots, \cellprog_n)$), or a filter construct of the form $\filterconstruct(\cellprog_1, \cellprog_2, \cellprog_3, \lambda \startcell. \lambda \itercell. \pred)$. Here, $\tau$ denotes a so-called \emph{cell program} for extracting a \emph{single} cell. The semantics of the $\filterconstruct$ construct is that it returns all cells between $\cellprog_2$ and $\cellprog_3$ that satisfy the predicate $\phi$. Note that the predicate $\phi$ takes two arguments $y$ and $z$, where $y$ is bound to the result of $\tau_1$ and $z$ is bound to each of the cells between $\tau_2$ and $\tau_3$.

The key building block of cell extraction programs is the $\getcell$ construct. In the simplest case, a $\getcell$ construct has the form $\getcell(x, \dir, k,  \lambda \startcell. \lambda \itercell. \pred)$ where $x$ is a cell, $\dir$ is a direction (up \updir, down \downdir, left \leftdir, right \rightdir), $k $ is an integer constant drawn from the set $ \{ -3, -2, -1, 1, 2, 3 \}$, and $\pred$ is a predicate. The semantics of this construct is that it finds the $k$'th cell satisfying predicate $\phi$ in direction $\dir$ from the starting cell $x$. For instance, the expression $\getcell(x, \rightdir, 1,  \lambda \startcell. \lambda \itercell. \true)$ refers to $x$ itself, while $\getcell(x, \rightdir, 2,  \lambda \startcell. \lambda \itercell. \true)$ extracts the neighboring cell to the right of cell $x$. An interesting point about the $\getcell$ construct is that it is recursive: For instance,  if $x$ is bound to cell $(r, c)$, then the expression
\[
\getcell(\getcell( x, \updir, 2, \lambda \startcell. \lambda \itercell. \true ), \rightdir, 2,  \lambda \startcell. \lambda \itercell. \true)
\]
retrieves the cell at row $r - 1$ and column $c+1$. Effectively, the recursive $\getcell$ construct allows the program to ``make turns" when searching for the target cell. As we observed in \exref{maketurn} from Section~\ref{sec:examples}, the ability to ``make turns" is necessary for expressing many real-world data extraction tasks.

\begin{figure*}[!t]
\small
\begin{minipage}{0.58\linewidth}
\[
\begin{aligned}
\semantics{\t{Val}(\mapper(\itercell)) = s}_{\inputtable, \constcell_1, \constcell_2} & = \text{Eval}(\inputtable(\mapper(\constcell_2)), =, s) \\
\semantics{\t{Val}(\mapper(\itercell)) \neq s}_{\inputtable, \constcell_1, \constcell_2} & = \text{Eval}(\inputtable(\mapper(\constcell_2)) , \neq, s) \\
\semantics{\valeq{\mapper(\startcell)}{\mapper(\itercell)}}_{\inputtable, \constcell_1, \constcell_2} & = \text{Eval}(\inputtable(\mapper(\constcell_1)), =, \inputtable(\mapper(\constcell_2)) ) \\ 
\semantics{\pred_1 \conj \pred_2}_{\inputtable, \constcell_1, \constcell_2} & = \semantics{\pred_1}_{\inputtable, \constcell_1, \constcell_2} \conj \semantics{\pred_2}_{\inputtable, \constcell_1, \constcell_2} \\ 
\semantics{x}_{\inputtable, \constcell} & = c \\ 
\semantics{\getcell(\cellprog, \dir, k, \lambda\startcell.\lambda\itercell.\pred)}_{\inputtable, \constcell} & = \left\{
\begin{aligned}
& \bottom \tab\tab\tab \langif ~ \semantics{\cellprog}_{\inputtable, \constcell} = \bottom  \text{ or } |k| > \emph{len}(L) \\ 
& L.\emph{get}\big{(} k \big{)} \tab\tab\tab\tab\tab\tab\tab~~~   \langif ~ k > 0  \\  
& L.\emph{get}\big{(} \emph{len}(L) + 1 - |k| \big{)} \tab ~~ \langif ~ k < 0  \\  
\end{aligned} 
\right. \\
& \tab  \text{where } L = \texttt{filter} \big{(} \texttt{range}(\semantics{\cellprog}_{\inputtable, \constcell}, \dir), (\lambda\startcell.\lambda\itercell. ~ \pred) \semantics{\cellprog}_{\inputtable, \constcell} \big{)} \\
\semantics{\unionconstruct(\cellprog_1, \dots, \cellprog_n)}_{\inputtable, \constcell} & = \lift{\semantics{\cellprog_1}}_{\inputtable, \constcell} \starunion \dots \starunion \lift{\semantics{\cellprog_n}}_{\inputtable, \constcell} \\ 
\semantics{\filterconstruct(\cellprog_1, \cellprog_2, \cellprog_3, \lambda\startcell.\lambda\itercell.\pred)}_{\inputtable, \constcell} & = \left\{
\begin{aligned}
& \bottom \tab\tab\tab\tab\tab\tab\tab ~~ \langif ~ \semantics{\cellprog_1}_{\inputtable, \constcell} = \bottom \text{ or } \semantics{\cellprog_2}_{\inputtable, \constcell} = \bottom \text{ or } \semantics{\cellprog_3}_{\inputtable, \constcell} = \bottom \\
& \texttt{filter} \big{(} \texttt{range}(\semantics{\cellprog_2}_{\inputtable, \constcell}, \semantics{\cellprog_3}_{\inputtable, \constcell}), (\lambda\startcell\lambda\itercell. ~ \pred) \semantics{\cellprog_1}_{\inputtable, \constcell} \big{)} \tab\tab \text{otherwise} \\  
\end{aligned}
\right. \\ 
\semantics{\seqconstruct(\simpleprog, \seqprog)}_{\inputtable, \constcell} & = \left\{
\begin{aligned}
\semantics{\simpleprog}_{\inputtable, \constcell} \tab & \langif ~ \semantics{\simpleprog}_{\inputtable, \constcell} \neq \bottom \\
\semantics{\seqprog}_{\inputtable, \constcell} \tab & \text{otherwise}
\end{aligned} 
\right.
\\ 
\end{aligned}
\]
\end{minipage}
%\vrule
\begin{minipage}{0.4\linewidth}
\[
\begin{aligned}
\text{Eval}(s_1, \triangleleft, s_2) & = \left\{
\begin{aligned}
& \text{false} & \langif ~ s_1 = \missingval \text{ or } s_2 = \missingval \\ 
& s_1 \triangleleft s_2 & \text{otherwise} \\ 
\end{aligned}
\right. \\ \\
\lift{\constcell} & = \left\{
\begin{aligned}
& \bottom \tab & \langif ~ \constcell = \bottom \\ 
& [ \constcell ] \tab & \text{otherwise} \\ 
\end{aligned}
\right. \\ 
\lift{\constcell_1} \starunion \lift{\constcell_2} & = \left\{
\begin{aligned}
& \bottom \tab & \langif ~ \lift{\constcell_1} = \bottom \text{or} ~ \lift{\constcell_2} = \bottom \\ 
& \lift{\constcell_1} :: \lift{\constcell_2} \tab & \text{otherwise} \\ 
\end{aligned}
\right. \\ \\ \\ \\ \\ \\ \\ 
\end{aligned}
\]
\end{minipage}
\caption{Semantics of the $\dsl$. $\inputtable(\cdot)$ gives the value of the given cell in table $\inputtable$. Function $\texttt{range}(\constcell, \dir)$ returns all  cells in the direction $\dir$ from $\constcell$ (including $\constcell$). Function $\texttt{range}(\constcell_1, \constcell_2)$ returns all  cells between $\constcell_1$ and $\constcell_2$ (including $\constcell_1$ and $\constcell_2$) if $\constcell_1$ and $\constcell_2$ are in the same row/column, otherwise it gets stuck. Function $\texttt{filter}(L, p)$ is the standard combinator that returns a list of elements satisfying $p$ in $L$.}
\figlabel{langsemantics}
\vspace{-0.1in}
\end{figure*}

Another important point about the $\getcell$ construct is that it returns $\bot$ if the $k$'th entry from the starting cell falls outside the range of the table. For instance, if the input table has $3$ rows, then  $\getcell( (3, 1), \downdir, 2, \lambda \startcell. \lambda \itercell. \true )$ yields $\bot$. Finally, another subtlety about $\getcell$ is that the $k$ value can be negative. For instance, $\getcell( x, \updir, -1,$ $\lambda \startcell. \lambda \itercell. \true )$ returns the uppermost cell in $x$'s column.

So far, we have seen how the $\getcell$ construct allows us to express spatial (geometrical) relationships by specifying a direction and a distance. However, as illustrated  through the examples in Section~\ref{sec:examples}, many real-world data extraction tasks require combining geometrical and relational reasoning. For this purpose, predicates in our DSL can be constructed using {conjunctions} of an expressive family of relations. Specifically, unary predicates $\valeqconst{\mapper(\itercell)}{s}$ and $\valneqconst{\mapper(\itercell)}{s}$ in our DSL check whether or not the value of a cell $\mapper(\itercell)$ is equal to a string constant $s$. Similarly, binary predicates $\valeq{\mapper(\startcell)}{\mapper(\itercell)}$ check whether two cells contain the same value.
Observe that the mapper function $\chi$ used in the predicate yields a new cell that shares some property with its input cell $\itercell$. For instance, the cell mapper $\colmapper{1}$ yields a cell that has the same row as $c$ but whose column is $1$. The use of mapper functions in predicates allows us to further combine geometric and relational reasoning.

\begin{example}
\figref{codeforexamples} gives the desired cell extraction program for each hole from \figref{examplespecs}. For instance, the program \textnormal{$\lambda\inputtable.\lambda\inputcell. ~ \getcell(\inputcell, \leftdir, 1, \lambda\startcell.\lambda\itercell. \valneqconst{\itercell}{\missingval})$} yields the first non-missing value to the left of $x$. Similarly, the cell extraction program for Example 2 yields the first cell $c$ such that (1) $c$ does not have a missing value, (2) $c$ has the same entry as $x$ at column $1$, and (3) $c$ is obtained by going up from $x$.
\end{example}

\begin{figure*}[!t]
\small
\centering
\begin{tabular}{ | c | c || m{11.6cm} | }
\hline
{\bf Task} & {\bf Sketch} & \multicolumn{1}{>{\centering\arraybackslash}m{11.6cm}|}{\bf Implementation in our DSL} \\ \hline 
%Figure 1 & $\hole_1 \holemapsto \lambda\inputtable.\lambda\inputcell. ~ \getcell(x, \leftdir, 1, \lambda\startcell.\lambda\itercell. \valneqconst{\itercell}{\texttt{WS}}), \hole_2 \holemapsto \lambda\inputtable.\lambda\inputcell. ~ \getcell(\inputcell, \leftdir, \text{-}2, \lambda\startcell.\lambda\itercell. \valneqconst{\itercell}{\texttt{WS}})$ \\ \hline 
Example 2.1 & $\texttt{SUM}(\sketchhole_1, 1)$ & $\sketchhole_1 \holemapsto \lambda\inputtable.\lambda\inputcell. ~ \getcell(\inputcell, \leftdir, 1, \lambda\startcell.\lambda\itercell. \valneqconst{\itercell}{\missingval})$ \\ \hline 
Example 2.2 & $\sketchhole_1$ & $\sketchhole_1 \holemapsto \lambda\inputtable.\lambda\inputcell. ~ \getcell \big{(} \inputcell, \updir, 1, \lambda\startcell.\lambda\itercell. \valneqconst{\itercell}{\missingval} \conj \valeq{(\fst(\startcell), 1)}{(\fst(\itercell), 1)} \big{)}$ \\ \hline 
Example 2.3 & $\sketchhole_1$ & $\sketchhole_1 \holemapsto \lambda\inputtable.\lambda\inputcell. ~ \getcell \big{(} \getcell( \getcell(\inputcell, \leftdir, 2, \lambda\startcell.\lambda\itercell.\true), \updir, 1, \lambda\startcell.\lambda\itercell.\valeqconst{(\fst(\itercell), 2)}{``1"}),$ 
$\downdir, 1, \lambda\startcell.\lambda\itercell.\valneqconst{\itercell}{``0"} \big{)}$ \\ \hline 
%Example 4 & $\texttt{AVG}(\hole_1, \hole_2)$ & \specialcell{ 
%$\hole_1 \holemapsto \lambda\inputtable.\lambda\inputcell. \getcell \big{(} \inputcell, \updir, 1, \lambda\startcell.%\lambda\itercell. \valneqconst{\itercell}{\missing} \big{)}$ 
%\\ 
%$\hole_2 \holemapsto \lambda\inputtable.\lambda\inputcell. \getcell \big{(} \inputcell, \downdir, 1, \lambda\startcell.\lambda\itercell. \valneqconst{\itercell}{\missing} \big{)} $} \\ \hline 
Example 2.4 & $\texttt{COUNT}(\sketchhole_1)$ & $\sketchhole_1 \holemapsto \lambda\inputtable.\lambda\inputcell. ~ \filterconstruct(\inputcell, \getcell(\inputcell, \updir, 1, \valeqconst{\itercell}{``~"}), \inputcell, \lambda\startcell.\lambda\itercell. \valneqconst{\itercell}{``~"})$ \\ \hline 
Example 2.5 & $\sketchhole_1$ & $\sketchhole_1 \holemapsto \lambda\inputtable.\lambda\inputcell. ~ \seqconstruct \big{(} \getcell(\inputcell, \updir, 1, \lambda\startcell.\lambda\itercell. \valneqconst{\itercell}{\missingval}), \getcell(\inputcell, \downdir, 1, \lambda\startcell.\lambda\itercell. \valneqconst{\itercell}{\missingval}) \big{)}$ \\ \hline 
\end{tabular}
\vspace{-0.1in}
\caption{Code in the $\dsl$ for examples. The $\unionconstruct$ keyword is omitted if it has only one cell program as the argument.}
\figlabel{codeforexamples}
\vspace{-18pt}
\end{figure*}

\section{Top-Level Synthesis Algorithm}\label{sec:synthesis}

We now present the synthesis algorithm for learning cell extraction programs  from input-output examples. Recall that the formula sketches provided by the user can contain multiple holes; hence, the synthesis algorithm is invoked once per hole in the input sketch, given input-output examples for each hole.

\begin{figure}[!t]
\vspace{-0.1in}
\begin{algorithm}[H]
\begin{algorithmic}[1]
\algnewcommand\sub{\mathcal{P}}
\algnewcommand\Input{\textbf{input: }} 
\algnewcommand\Output{\textbf{output: }}
\Procedure{\learn}{$\text{table} ~ \inputtable, \text{set} ~ \examples$} \\ 
\vspace{5pt}
\Input{Table $\inputtable$, and input-output examples $\examples$ of the form $\{ i_1 \mapsto L_1, \dots, i_n \mapsto L_n \}$} \\ 
\Output{A $\seqconstruct$ program satisfying all examples in $\examples$} 
\vspace{5pt}
\State $\seqprog \assign \failure$;
\For{$k \assign 1, \dots, |\examples|$} \Comment{Enumerate \# branches}
\State $\seqprog \assign \textsc{\learnseq}(\inputtable, k, \examples)$; 
\If{$\seqprog \neq \failure$} \Return{$\seqprog$;} \EndIf
%\If{$\score(\seqprog) > \score(\seqprog^\prime)$} $\seqprog^\prime \assign \seqprog$; \EndIf
\EndFor
%\State\Return{$\seqprog^\prime$;}
\State\Return{$\failure$;}
\EndProcedure
\end{algorithmic}
\end{algorithm}
\vspace{-36pt}
\caption{Algorithm for $\textsc{\learn}(\inputtable, \mathcal{E})$.}
\figlabel{learn}
\end{figure}

\figref{learn} shows the high-level structure of our synthesis algorithm. 
The algorithm {\sc Learn} takes as input a table $\inputtable$ and a set of examples $\examples$ of the form $\{ i_1 \mapsto L_1, \dots, i_n \mapsto L_n \}$ and returns a $\seqconstruct$ program $\pi$ such that  $\pi$ is consistent with all examples $\ex$. Essentially, the {\sc Learn} algorithm enumerates the number of branches $k$ in the $\seqconstruct$ program $\seqprog$. 
In each iteration of the loop (lines 5--7), it calls the {\sc LearnExtractor} function to find the ``best" program that contains exactly $k$ branches and satisfies all the input-output examples, and returns the program with the minimum number of branches. 
If no such program is found after the loop exits, it returns $\failure$, meaning there is no program in the $\dsl$ satisfying all examples in $\examples$.~\footnote{A discussion of the complexity of the algorithm can be found in the Appendix.}
%The return value of {\sc Learn} is the program that has the highest score according to a heuristic score function $\theta$.~\footnote{Note that {\sc LearnExtractor} can return $\failure$, meaning that there does not exist a program in the $\dsl$ with the specified number of branches. We assume $\failure$ is assigned the lowest score, i.e., $\theta(\failure) = -\infty$.}
%learns the best $\seqconstruct$ program. It finally returns the best program among programs with different number branches. 

Let us now consider the {\sc LearnExtractor} procedure in \figref{learnseq} for learning a program with exactly $k$ branches. At a high level, {\sc LearnExtractor} partitions the inputs $\ex$ into $k$ disjoint subsets $\mathcal{S}_1, \dots, \mathcal{S}_k$ and checks whether each $\mathcal{S}_i$ can be ``unified", meaning that all examples in  $\mathcal{S}_i$   can be represented using the same conditional-free program. Our algorithm performs unification  using the {\sc LearnSimpProg} function, which we will explain later.

Let us now consider the recursive {\sc LearnExtractor} procedure in a bit more detail. The base case of this procedure corresponds to learning a simple (i.e., conditional-free) program. Since the synthesized program should not introduce any branches, it simply calls the {\sc LearnSimpProg} function to find a unifier $\rho$ for examples $\ex$. Note that the conditional-free program $\simpleprog$ here could be $\failure$.  
%If {\sc LearnSimpProg} returns $\failure$, there is no conditional-free program satisfying $\ex$; hence the algorithm returns $\failure$ (line 6) \todo{Do we really need line 6?}. 
%Otherwise, we return the program $\seqconstruct(\unionprog)$.

\begin{figure}[!t]
\begin{algorithm}[H]
\begin{algorithmic}[1]
\algnewcommand\examples{\mathcal{E}}
\algnewcommand\sub{\mathcal{P}}
\algnewcommand\Input{\textbf{input: }} 
\algnewcommand\Output{\textbf{output: }}
\Procedure{\learnseq}{$\text{table} ~ \inputtable, \text{int} ~ k, \text{set} ~ \examples$} \\ 
\vspace{5pt}
\Input{Table $\inputtable$, number of partitions $k$, and examples $\examples$ of the form $\{ i_1 \mapsto L_1, \dots, i_n \mapsto L_n \}$} \\ 
\Output{A $\seqconstruct$ program with $k$ branches satisfying $\examples$} 
\vspace{5pt}
\If{$k = 1$}
\Comment{Base case}
\State $\unionprog \assign \textsc{\learnsimple}(\inputtable, \examples, \emptyset)$;
%\If{$~ \unionprog = \failure$} \Return{$\failure$;} \EndIf
\State \Return{$\unionprog$;}
\EndIf
\vspace{5pt}
\State $\seqprog^\prime \assign \failure$;\Comment{Recursive case}
%\ForAll{$\sub \in 2^{\t{\dom}(\examples)} \setminus \{ \t{\dom}(\examples), \emptyset \}$}
\ForAll{$\sub \in 2^{\t{\dom}(\examples)}$}
\State $\unionprog \assign \textsc{\learnsimple}(\inputtable, \examples \project \sub, \overline{\sub})$;
\If{$\unionprog = \failure$} \continue; \EndIf
\State $\seqprog \assign \textsc{\learnseq}(\inputtable, k - 1, \examples \project \overline{\sub})$;
\If{$\seqprog = \failure$} \continue; \EndIf
\State $\seqprog^{\prime\prime} \assign \seqconstruct(\unionprog, \seqprog)$;
\If{$\score(\seqprog^{\prime\prime}) > \score(\seqprog^\prime)$} $\seqprog^\prime \assign \seqprog^{\prime\prime}$; \EndIf
\EndFor
\State \Return{$\seqprog^{\prime}$;}
\EndProcedure
\end{algorithmic}
\end{algorithm}
\vspace{-30pt}
\caption{Algorithm for $\textsc{\learnseq}(\inputtable, k, \mathcal{E})$.}
\figlabel{learnseq}
\end{figure}

The recursive case of {\sc LearnExtractor} enumerates all subsets of all the input cells $\t{\dom}(\examples)$ that are to be handled by the first simple program in the $\seqconstruct$ construct (line $8$). Here the notation $\t{\dom}(\examples)$ is defined as follows: 
\[
\t{\dom}(\examples) = \{ i \ | \ (i \mapsto L) \in \examples \}
\]
Given a subset $\mathcal{P}$ of all the input cells, we first try to unify the examples whose inputs belong to $\mathcal{P}$. In particular, the notation  $\ex \downarrow \mathcal{P}$ used at line $9$ is defined as follows:
\[
\ex \downarrow \mathcal{P} = \{ i \mapsto L \ | \ i \in \mathcal{P} \land (i \mapsto L) \in \ex \}
\]
Hence, if the call to {\sc LearnSimpProg} at line $9$ returns non-null, this means  there is a conditional-free program $\rho$ satisfying  all examples in $\ex \downarrow \mathcal{P}$ and returning failure ($\bot$) on all input cells $\overline{\mathcal{P}} = \t{\dom}(\ex) \backslash \mathcal{P}$. 
Note that it is crucial that $\rho$ yields $\bot$ on inputs $\overline{\mathcal{P}}$, because the second statement of the $\seqconstruct$ construct is only evaluated when the first program returns $\bot$. For this reason, the unification algorithm {\sc LearnSimpleProg} takes as input a set of negative examples $\ex^{-}$ as well as positive examples $\ex^+$. Observe that a ``negative example" in our case corresponds to a cell on which the learned program should yield $\bot$.

Now, if unification of examples $\ex \downarrow \mathcal{P}$ returns $\failure$, we know that $\mathcal{P}$ is not a valid subset and we move on to the next subset. On the other hand, if $\ex \downarrow \mathcal{P}$ is unifiable, we  try to construct a program that contains $k-1$ branches and that is consistent with the remaining examples $\ex \downarrow \overline{\mathcal{P}}$. If such a program exists (i.e, the recursive call to {\sc LearnExtractor} at line $11$ returns non-null), we have found a program $\seqconstruct(\rho, \pi)$ that has exactly $k$ branches and that satisfies all input-output examples $\ex$. However, our algorithm does not return the first consistent program it finds: In general, our goal is to find the ``best"  program, according to some heuristic cost metric $\theta$ (described in the Appendix), that is consistent with the examples and that minimizes the number of branches. For this reason,  the algorithm only returns $\pi^\prime$ after it has explored all possible partitions (line $15$).

%The algorithm also maintains and updates the ``best" $\seqconstruct$ program $\pi^\prime$ it has explored according to a score function $\score$~\footnote{We will discuss the design of the score function $\score$ in Section 6.3.} (line $14$), and it returns $\pi^\prime$ after it has explored all possible partitions (line $15$).

\begin{figure}[!t]
%\vspace*{-2pt}
\begin{algorithm}[H]
\begin{algorithmic}[1]
\algnewcommand\examples{\mathcal{E}}
\algnewcommand\posexs{\examples^{+}}
\algnewcommand\negexs{\examples^{-}}
\algnewcommand\progset{\mathcal{S}}
\algnewcommand\Input{\textbf{input: }} 
\algnewcommand\Output{\textbf{output: }}
\Procedure{\learnsimple}{$\text{table} ~ \inputtable, \text{set} ~ \posexs, \text{set} ~ \negexs$} \\ 
\vspace{5pt}
\Input{ $\posexs$ is positive examples $\{ i_1 \mapsto L_1, \dots, i_n \mapsto L_n \}$, and  $\negexs: \{ j_1, \dots, j_m \}$ is set of negative examples} \\ 
\Output{A simple program satisfying both $\posexs$ and $\negexs$} 
\vspace{5pt}
%\State $\simpleprog \assign \texttt{read\_cache}(\posexs, \negexs)$; \Comment{Read cache}
%\If{$\simpleprog \neq \failure$} \Return{$\simpleprog$} \EndIf
%\vspace{5pt}
\State \Comment{Initialization}
\If{$\exists i, i^\prime \in \textnormal{dom}(\posexs). \ \emph{len}(\posexs[i]) \neq \emph{len}(\posexs[i^\prime]) $ 
\State} $t \assign -1$;
\Else{\ $t \assign \emph{len}(\posexs[i_1])$;}
\EndIf
\State $\progset \assign \textsc{\buildprogset}(\inputtable, i_1, \posexs[i_1], t)$;
%\State $\progset \assign \failure$;
\vspace{0.05in}
\For{$i \in \textnormal{\dom}(\posexs) \setminus \{ i_1 \}$} \Comment{Positive examples}
\State $\progset \assign \progset ~ \intersect ~ \textsc{\buildprogset}(\inputtable, i, \posexs[i], t)$;
%\If{$\progset = \failure$} $\progset \assign \textsc{\buildprogset}(\inputtable, i, \posexs[i], n)$;
%\Else{\ $\progset \assign \progset ~ \intersect ~ \textsc{\buildprogset}(\inputtable, i, \posexs[i], n)$;}
%\EndIf
\EndFor
\vspace{0.05in}
\For{$j \in \negexs$} \Comment{Negative examples}
\State $\progset \assign \progset ~ \intersect ~ \textsc{\buildprogset}(\inputtable, j, \bottom, t)$;
\EndFor
\If{$\semantics{\progset} = \emptyset$} \Return{$\failure$;} \EndIf
\State $\unionprog \assign \textsc{\rank}(\progset)$; \Comment{Rank programs}
%\vspace{5pt}
%\State $\texttt{store\_cache}(\posexs, \negexs, \simpleprog)$; \Comment{Store cache}
%\vspace{5pt}
\State \Return{$\unionprog$;}
\EndProcedure 
\end{algorithmic}
\end{algorithm}
\vspace{-30pt}
\caption{Algorithm for $\textsc{\learnsimple}(\inputtable, \mathcal{E}^{+}, \mathcal{E}^{-})$.}
\figlabel{learnsimple}
\vspace{-4pt}
\end{figure}

Finally, let us turn our attention to the unification algorithm {\sc LearnSimpProg} from \figref{learnsimple} for finding a simple (conditional-free) program $\rho$ such that $\rho(\inputtable, i) = L$ for all $(i \mapsto L) \in \ex^+$ and $\rho(\inputtable, i) = \bot$ for every $i \in \ex^-$.
The key idea underlying {\sc LearnSimpProg} is to construct a finite tree automaton $\mathcal{A}$ for each example $i \mapsto L$ such that $\mathcal{A}$ accepts \emph{exactly} those simple programs that produce $L$ on input $i$.~\footnote{Note that here $L$ can be either a list of cells or $\bottom$.} Since constructing the FTA for a given example is non-trivial, we will explain it in detail in the next section. Once we construct an FTA for each example, the final program is obtained by taking the intersection of the tree automata for each individual example (lines $8$-$12$) and then extracting the best program accepted by the automaton (line $14$).

\begin{example}
Consider a row vector \textnormal{$[ \missingval, 1, \missingval, 2, \missingval ]$}. The user wants to fill each missing value by the \emph{previous} non-missing value, if one exists, and by the \emph{next} non-missing value otherwise. Here, the formula sketch is \textnormal{$\sketchhole_1$}, and suppose that the user gives two examples for \textnormal{$\sketchhole_1$}: 
\[ (1,1) \mapsto (1,2) \ \ \emph{and} \ \  (1,3) \mapsto (1,2) \]
Given this input, \tool first tries to learn a \textnormal{$\seqconstruct$} program with one branch by unifying two examples.  Since there is no such program in the $\dsl$, the algorithm then tries to learn a  program \textnormal{$\seqprog = \seqconstruct(\simpleprog_1, \simpleprog_2)$}  by partitioning the examples into two disjoint sets. There are two ways of partitioning the examples: 
\begin{enumerate}
\item  The first partition is $\{(1,1)\}$, meaning that $\simpleprog_1$ should return $(1,2)$ on $(1,1)$ and $\bot$ on $(1, 3)$, and $\simpleprog_2$ should return $(1,2)$ on $(1, 3)$. In this case, \tool learns the following two programs $\simpleprog_1, \simpleprog_2$:
\textnormal{
\[
\begin{array}{ll}
\simpleprog_1: & \getcell(\inputcell, \rightdir, -2, \lambda\startcell.\lambda\itercell. \valneqconst{\itercell}{\missingval}) \\
\simpleprog_2: & \getcell(\inputcell, \leftdir, 1, \lambda\startcell.\lambda\itercell. \valneqconst{\itercell}{\missingval})
\end{array}
\]
}
\item  The second partition is $\{(1,3)\}$, meaning that $\simpleprog_1$ should return $(1,2)$ on $(1,3)$ and $\bot$ on $(1, 1)$, and $\simpleprog_2$ should return $(1,2)$ on $(1, 1)$. In this case, \tool learns the following two programs $\simpleprog_1, \simpleprog_2$:
\textnormal{
\[
\begin{array}{ll}
\simpleprog_1: & \getcell(\inputcell, \leftdir, 1, \lambda\startcell.\lambda\itercell. \valneqconst{\itercell}{\missingval}) \\
\simpleprog_2: & \getcell(\inputcell, \rightdir, 1, \lambda\startcell.\lambda\itercell. \valneqconst{\itercell}{\missingval})
\end{array}
\]
}
\end{enumerate} 
After learning these two programs, \tool compares them according to a scoring function $\theta$. In this case, the second program has a higher score (because the scoring function prefers $k$ values in \textnormal{$\getcell$} with smaller absolute value), hence we  select the \textnormal{$\seqconstruct$} program learned in the second case.
\exlabel{synthesisexample}
\end{example}

\section{Unification using Tree Automata}\label{sec:representation}

In this section, we describe how to represent conditional-free programs using \emph{finite tree automata (FTA)}.
Since the {\sc LearnSimpProg} algorithm from Section~\ref{sec:synthesis} performs unification using  standard FTA intersection~\cite{tata2007},  the key problem is how to construct a tree automaton representing all  programs that are consistent with a given input-output example. As mentioned in Section~\ref{sec:intro}, our FTA-based formulation leads to a new version space learning algorithm that offers several advantages over prior VSA-based techniques. We discuss the relationship between our learning algorithm and VSA-based techniques in Section~\ref{sec:vsa}.

%\xinyu{-- This needs to be consistent with the intro (Xinyu). As mentioned in Section~\ref{sec:intro}, we use FTA rather than version space algebra (VSA) for two reasons: 1) FTA yields a more compact representation for our DSL, and 2) it is non-trivial to encode the inverse semantics of constructs in our DSL.}
%In the case of our DSL, observe that the operands of the $\filterconstruct$ construct are dependent: For instance, it is not possible learn the start and end cells $\cellprog_2, \cellprog_3$ in the $\filterconstruct$ construct independently from the predicate $\pred$. 

\subsection{Preliminaries}

Since the remainder of this section relies on basic knowledge about finite tree automata, we first provide some background on this topic. At a  high level, tree automata generalize standard automata by accepting trees rather than words (i.e., chains).

\begin{definition}[FTA]
A (bottom-up) finite tree automaton (FTA) over alphabet $\alphabet$ is a tuple $\mathcal{A} = (\states, \alphabet, \finalstates, \transitionrules)$ where $\states$ is a set of states, $\finalstates \subseteq \states$ is a set of final states, and $\transitionrules$ is a set of transition (rewrite) rules  of the form
$f(q_1, \dots, q_n) \rewriteto q$
where $q,q_1,\dots,q_n \in \states$, and $f \in \alphabet$ has arity $n$.
\end{definition}

\begin{wrapfigure}{h}{0.35\linewidth}
\vspace{-0.3in}
\begin{center}
\includegraphics[scale=0.45]{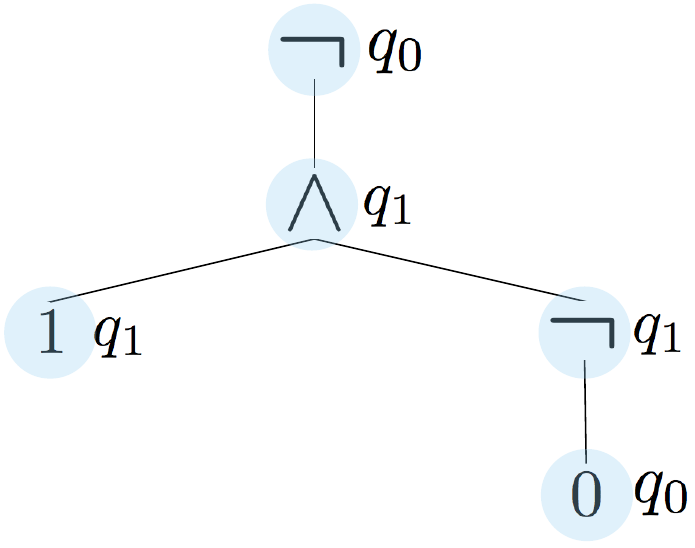}
\end{center}
\vspace{-0.1in}
\caption{Tree for $\neg (1 \land \neg 0)$ where each sub-term is annotated with its state.}
\figlabel{extree}
\vspace{-0.1in}
\end{wrapfigure}

Here, the alphabet is \emph{ranked} in that every symbol $f$ in $\alphabet$ has an arity (rank) associated with it. The set of function symbols of arity $k$  is denoted  $\alphabet_k$. In the context of tree automata, we view ground terms over alphabet $\alphabet$ as trees. Intuitively, an FTA accepts a ground term $t$ if we can rewrite $t$ to some $q \in \finalstates$ using rules in $\transitionrules$. The language of an FTA $\automaton$, written $\lang(\automaton)$, is the set of all ground terms accepted by $\automaton$.

\begin{example}
Consider the tree automaton defined by states $Q = \{ q_0, q_1 \}$, $\alphabet_0 = \{ 0, 1 \}$, $\alphabet_1 = \{ \neg \}$ $\alphabet_2 = \{ \land \}$, final state $q_1$, and the following transitions $\Delta$:
\[
\hspace*{-2in}
\begin{array}{ l l l l }
1 \rightarrow q_1 & 0 \rightarrow q_0 & \land(q_0, q_0) \rightarrow q_0 & \land(q_0, q_1)  \rightarrow q_0 \\
\neg(q_1) \rightarrow q_0 & \neg(q_0) \rightarrow q_1 & \land(q_1, q_0) \rightarrow q_0 & \land(q_1, q_1)  \rightarrow q_1 \\
\end{array}
\]
This tree automaton accepts exactly those propositional logic formulas (without variables) that evaluate to \emph{true}. 
As an example, \figref{extree} shows the tree for the formula $\neg (1 \land \neg 0)$ where each sub-term is annotated with its state. Since $q_0$ is not a final state, this formula is rejected by the tree automaton.
%As we can see, this formula is rewritten to the state $q_0$, therefore it is not accepted by the tree automaton. 
\end{example}

\subsection{Simple programs as tree automata}

Suppose we are given a single input-output example $i \mapsto L$ where we have $i \in \cells(\inputtable)$ and $L$ is either $\bot$ or a list of cells $[o_1, \dots, o_t]$ in table $\inputtable$. Our key idea is to construct a (deterministic) bottom-up FTA $\automaton$ such that $\lang(\automaton)$ represents exactly those simple programs (i.e., \emph{abstract syntax trees}) that produce output $L$ on input cell $i$.

Before presenting the full FTA construction procedure, we first explain the intuition underlying our automata. At a high level, the alphabet of the FTA corresponds to our DSL constructs, and the states in the FTA represent cells in table $\inputtable$. The constructed FTA  contains  a transition  $f(c_1, \dots, c_n) \rightarrow c$ if it is possible to get to cell $c$ from cells $c_1, \dots c_n$ via the DSL construct $f$. Hence, given an input cell $c$ and output cell $c'$, the trees accepted by our FTA correspond to simple programs that produce output cell $c'$ from input cell $c$.

\begin{figure}[!t]
\small 
\[
\begin{array}{ c c c c }
\irule{
}{
\lambda c. c \in \mappers(\inputtable) 
}
& \tab
\irule{
\begin{array}{c}
1 \leq k \leq \t{NumCols}(\inputtable) \\ 
\end{array}
}{
\lambda c. (k, \snd(c)) \in \mappers(\inputtable)
}
& \tab
\irule{
\begin{array}{c}
1 \leq k \leq \t{NumRows}(\inputtable) \\ 
\end{array}
}{
\lambda c. (\fst(c), k) \in \mappers(\inputtable) 
}
& \tab
\irule{
}{
\true \in \preds(\inputtable)
}
\end{array}
\]
\vspace{10pt}
\[
\begin{array}{ c c c }
\irule{
\begin{array}{c}
\mapper \in \mappers(\inputtable), \ \ \triangleleft \in \{ =, \neq \} \\ 
\exists c \in \cells(\inputtable). \ \  s = \tvalue(c) \\ 
\end{array}
}{
 (\t{Val}({\mapper(z)}) \triangleleft s) \in \preds(\inputtable)
}
& \tab
\irule{
\begin{array}{c}
\mapper \in \mappers(\inputtable) \\ 
\end{array}
}{
\valeq{\mapper(y)}{\mapper(z)} \in \preds(\inputtable) 
}
& \tab
\irule{
\pred_1 \in \preds(\inputtable), \pred_2 \in \preds(\inputtable)
}{
\pred_1 \conj \pred_2 \in \preds(\inputtable)
}
\end{array}
\]
\caption{Rules for constructing the universe of predicates.}
\figlabel{predrules}
\end{figure}

With this intuition in mind, let us now consider the FTA construction procedure in more detail. Given a table $\inputtable$, an input cell $i$, a list of output cells  $L$ (or $\bot$ in the case of negative examples), and an integer $t$ denoting the number of output cells, we construct a tree automaton $\automaton = (Q, \alphabet, Q_f, \Delta)$ in the following way:
\begin{itemize}
\item The states $Q$ include all cells in table $\inputtable$ as well as two special symbols $q_\bot$ and $q^\star$:
\[  Q = \{ q_c \ | \ c \in \cells(\inputtable)\} \cup \{ q_\bot, q^\star \} \]
Here, $q_\bot$ denotes any cell that is outside the range of the input table, and $q^\star$ indicates we have reached all desired output cells.
\item The final states $Q_f$ only include the special symbol $q^\star$ (i.e., 
$Q_f = \{ q^\star \}$).
\item The alphabet $\alphabet$ is $\alphabet_0 \cup \alphabet_1 \cup \alphabet_3 \cup \alphabet_t$ where $\alphabet_0 = \{ x \}$, $\alphabet_t = \{ \unionconstruct \}$~\footnote{If $t$ is $-1$, then we have $\alphabet = \alphabet_0 \cup  \alphabet_1 \cup \alphabet_3$. Recall from \figref{learnsimple} that $t=-1$ indicates that the positive examples differ in the  number of output cells; hence, there can be no simple program constructed using \unionconstruct\ that is consistent with all input-output examples. }, and
\[
\alphabet_3 = \{ \filterconstruct_\phi \ | \ \phi \in \predicates(\inputtable)\}
\]
\[
\alphabet_1 = \left \{ \getcell_{\dir, k, \pred}  \ {\Big \vert }
\begin{array}{c}
\ \dir \in \{ \updir, \downdir, \leftdir, \rightdir \} \land 
 k \in [1,3] \cup [-3,-1] 
  \land \ \phi \in \predicates(\inputtable) 
\end{array}
\right
\}
\]
In other words, the alphabet of $\automaton$ consists of the DSL constructs for simple programs. Since  $\filterconstruct$ and $\getcell$ statements also use predicates, we construct the universe of predicates $\predicates(\inputtable)$ as shown in \figref{predrules} and generate a different symbol for each different predicate. 
Furthermore, since $\getcell$ also takes a direction and position as input, we also instantiate those arguments with concrete values.
\item The transitions $\Delta$ of $\automaton$ are constructed using the inference rules in \figref{transitionrules}. The {\rm Init} rule says that argument $x$  is bound to input cell $i$. The  {\rm GetCell} rule states that $\getcell_{\dir, k, \pred}(q_c)$ can be rewritten to $q_{c'}$ if we can get to cell $c'$ from $c$ using the $\getcell$ construct in the DSL. The {\rm List 1} rule says that we can reach the final state $q^\star$ if the arguments of $\unionconstruct$ correspond to cells in the output $L$. Finally, the {\rm Filter 1} rule states that we can reach the final state $q^\star$ from states $q_{c_1}, q_{c_2}, q_{c_3}$ via $\filterconstruct_{\phi}$ if $\filterconstruct(c_1, c_2, c_3, \phi)$ yields the output $L$ on input table $\inputtable$. 
The second variants of the rules (labeled 2) deal with the special case where $L = \bot$ (negative example). For instance, according to the List 2 rule, we can reach the final state $\finalstate$ if the desired output is $\bot$ and any of the arguments of $\unionconstruct$ is $\bot$. 
\end{itemize}

\begin{figure}[!t]
\small 
\[
\begin{array}{ c c }
\vspace{1pt}
\begin{array}{ c r }
\irule{
}{
\inputtable, i, L \vdash \inputcell \rewriteto \state_i \in \transitionrules 
} & {\rm (Init)} 
\\ \\ \\ 
\irule{
\begin{array}{c}
c \in \cells(\inputtable) \cup \{ \bot \} \\
\semantics{\getcell(c, \dir, k, \lambda\startcell.\lambda\itercell.\pred)}_{\inputtable} = c' \\
\end{array}
}{
\inputtable, i, L \vdash \getcell_{\dir, k, \pred}(\state_c) \rewriteto \state_{c'} \in \transitionrules
} & ({\rm GetCell})
\\ \\ 
\irule{
\begin{array}{c}
L = [ o_1, \dots, o_n ] \\
c_1 \in \cells(\inputtable), c_2 \in \cells(\inputtable), c_3 \in \cells(\inputtable) \\ 
\semantics{\filterconstruct(c_1, c_2, c_3, \lambda\startcell.\lambda\itercell.\pred)}_{\inputtable} = L \\ 
\end{array}
}{
\inputtable, i, L \vdash \filterconstruct_{\pred}(\state_{c_1}, \state_{c_2}, \state_{c_3}) \rewriteto \finalstate \in \transitionrules
} & ({\rm Filter} \ 1)
\end{array}
& 
\begin{array}{ c r }
\irule{
L = [o_1, \dots, o_n]
}{
\inputtable, i, L \vdash \unionconstruct(\state_{o_1}, \dots, \state_{o_n}) \rewriteto \finalstate \in \transitionrules 
}  & ({\rm List} \ 1)
\\ \\ 
\irule{
\begin{array}{c}
L = \bot \\
 \exists k \in [1, n]. \ \state_{j_k} = q_\bot \\
\forall k \in [1, n]. \ \state_{j_k} \neq q^\star
\end{array}
}{
\inputtable, i, L \vdash \unionconstruct(\state_{j_1}, \dots, \state_{j_n}) \rewriteto \finalstate \in \transitionrules 
} & ({\rm List} \ 2)
\\ \\ 
\irule{
\begin{array}{c}
L = \bot \\
 \exists k \in [1, 3]. \ \state_{j_k} = q_\bot \\
\forall k \in [1, 3]. \ \state_{j_k} \neq q^\star
\end{array}
}{
\inputtable, i, L \vdash \filterconstruct_{\pred}(\state_{j_1}, \state_{j_2}, \state_{j_3}) \rewriteto \finalstate \in \transitionrules
} & ({\rm Filter} \ 2) 
\end{array}
\end{array}
\]
\vspace{-0.1in}
\caption{Rules for constructing transition rules $\transitionrules$ for table $\inputtable$ and example $i \mapsto L$.} 
\figlabel{transitionrules} 
\vspace{-0.1in}
\end{figure}

\begin{theorem} {\bf (Soundness and Completeness)}
Let $\automaton$ be the finite tree automaton constructed by our technique for example $i \mapsto L$ and table \textnormal{$\inputtable$}. Then, $\automaton$ accepts the tree that represents a simple program $\simpleprog$ iff \textnormal{$\semantics{\simpleprog}_{\inputtable, i} = L$}.
\end{theorem}

\begin{proof}
\vspace{-0.1in}
Provided in the appendix.
\vspace{-0.1in}
\end{proof}

\begin{figure*}[!t]
\footnotesize
\centering
\begin{tabular}{ | >{\centering\arraybackslash} m{.2cm} || >{\centering\arraybackslash} m{3.8cm} | >{\centering\arraybackslash} m{4.6cm} | m{5.7cm} | }
\hline
\rot{Tree} & \vspace{3pt}\includegraphics[scale=0.48]{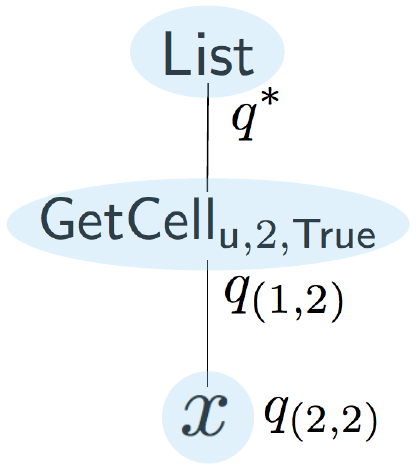} & \vspace{3pt}\includegraphics[scale=0.4]{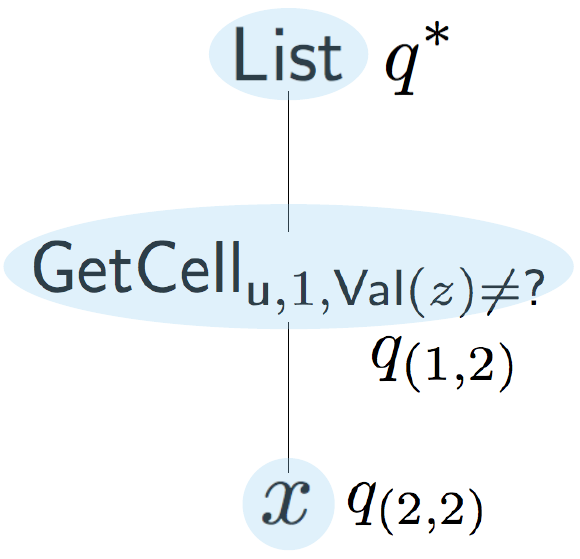} &
\multicolumn{1}{>{\centering\arraybackslash}m{5.7cm}|}{\vspace{3pt}\includegraphics[scale=0.47]{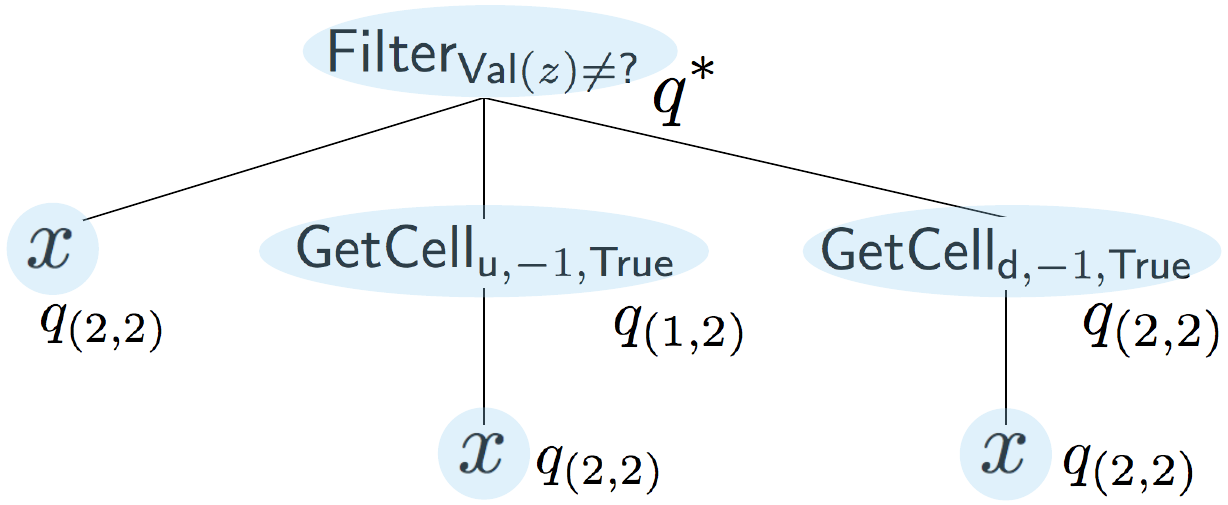}} \\ \hline
\rot{Prog.} & $\unionconstruct(\getcell(\inputcell, \updir, 2, \lambda\startcell.\lambda\itercell. \true))$ & $\unionconstruct(\getcell(\inputcell, \updir, 1, \lambda\startcell.\lambda\itercell.\valneqconst{\itercell}{\missingval}))$ & $\filterconstruct \big{(} \inputcell, \getcell(\inputcell, \updir, -1, \lambda\startcell.\lambda\itercell. \true), \getcell(\inputcell,$ $\tab\tab\tab \downdir, -1, \lambda\startcell.\lambda\itercell. \true), \lambda\startcell.\lambda\itercell. \valneqconst{\itercell}{\missingval} \big{)}$ \\ \hline 
\rot{Desc.} & The previous cell upwards & The previous non-missing cell upwards & \multicolumn{1}{>{\centering\arraybackslash}m{5.7cm}|}{All non-missing cells in the same column} \\ \hline 
\end{tabular}
\caption{Three terms (programs) accepted by the tree automaton in \exref{exfta}.}
\figlabel{exprogs}
\end{figure*}

\begin{example}
Consider a small table with two rows, where the first row is $[ 4, 5 ]$ and the second row is $[ 6, \missingval ]$. 
Furthermore, suppose the user provides the sketch \textnormal{$\sketchhole_1$} and the example $\{ (2,2) \mapsto [ (1,2) ] \}$ (i.e., $\missingval $ should be  $5$).

%Consider the small table shown in \figref{exfta}, where the missing value should be replaced by its previous non-missing value in the same column. 
%Furthermore, suppose the user provides the sketch \textnormal{$\hole_1$} and the example  $\{ (2,2) \mapsto [ (1,2) ] \}$. 

Let us consider the FTA construction for this example.
The states in the FTA are $ \{ \state_{(1,1)}, \state_{(1,2)}, \state_{(2,1)}, \state_{(2,2)}, \state_{\bottom}, \state^{*} \}$, 
and the final state is $ q^{*}$. 
The  transition rules $\transitionrules$ of the FTA are constructed according to \figref{transitionrules}. 
For example, $\transitionrules$ includes \textnormal{$\inputcell \rewriteto \state_{(2,2)}$} because cell $(2,2)$ is the input cell in the example \textnormal{(Init)}. It also includes \textnormal{$\getcell_{\updir, 2, \true}(\state_{(2,2)}) \rewriteto \state_{(1,2)}$} because, using {\rm GetCell}, we have \textnormal{
\[\semantics{\getcell((2,2), \updir, 2, \lambda\startcell.\lambda\itercell.\true)}_{\inputtable} = (1,2)\]} 
Using {\rm List 1}, the transition rules $\transitionrules$ also include the transition \textnormal{$\unionconstruct(\state_{(1,2)}) \rewriteto \state^{*}$}
and, using \textnormal{Filter 1}, it contains:
\textnormal{\[ \filterconstruct_{\valneqconst{\itercell}{\missingval}}(\state_{(2,2)}, \state_{(1,2)}, \state_{(2,2)}) \rewriteto \state^{*} \]}
\exlabel{exfta}
\end{example}

\vspace{-0.3in}
\subsection{Ranking programs accepted by tree automaton}
Recall from Section~\ref{sec:synthesis} that the {\sc LearnSimpProg} algorithm requires finding the ``best" program accepted by $\automaton$ according to a $\rank$ function. In this section, we briefly discuss how we rank programs accepted by a tree automaton. In the rest of this section, we use the term ``program" and its corresponding AST interchangably.

\begin{definition}
The \emph{size} of a tree (term) $t$, denoted by $|| t ||$, is inductively defined by: \\ 
\indent - $|| t || = 1$ \textnormal{if} $t \in \alphabet_0$, \\
\indent - $|| t || = 1 + \Sigma_{i \in \{ 1, \dots, n \}}{|| t_i ||}$ \textnormal{if} $\textnormal{Root}(t) \in \alphabet_n$, where $t_i$ is the $i$-th argument of $t$. 
\end{definition}
The ranking function $\rank(\progset)$ selects the best program by finding a tree $t$ in $\mathcal{L}(\progset)$ such that $|| t|| $ is minimized. If there are multiple trees with the same size, it selects the one with the highest score according to a heuristic scoring function $\score$. 
Intuitively, $\score$ assigns a positive score to each language construct in our DSL so that a more ``general'' construct has a higher score. 
For example, the identity mapper $\idmapper$ is assigned a higher score than the other cell mappers, and predicate $\true$ has a higher score than other predicates. More details about the heuristic scoring function $\score$ are provided in the appendix.

%The score of a term is computed by composing the scores of all its sub-terms in a heuristic way. For example, the score of a $\unionconstruct$ program is the weighted average of scores for its arguments. \xinyu{Please find more details about the design of $\score$ in appendix.}

\begin{example}
\figref{exprogs} shows a subset of the programs accepted by the tree automaton $\progset$ from \exref{exfta}. 
The first program, \textnormal{$\unionconstruct(\inputcell, \updir, 2, \lambda\startcell.\lambda\itercell.\true)$}, is selected as the best program, because it is of the minimum size (3), and it only uses the most general predicate \textnormal{$\true$}. 
\end{example}

\section{Implementation}

We have implemented the synthesis algorithm proposed in this paper in a tool called \tool, written in Java. While our implementation mostly follows our technical presentation, it performs several additional optimizations.
First, our presentation of {\sc LearnSimpProg} algorithm constructs a separate tree automaton for each example. However, observe that the tree automata for different examples actually have the same set of cells and share a large subset of the transitions. 
%For instance, all tree automata share the same set of $\getcell$ transitions, which typically constitute $99\%$ of all transitions. 
Based on this observation, our implementation constructs a base (incomplete) tree automaton $\automaton_B$ that is shared by all examples and then adds additional transitions to $\automaton_B$ for each individual example. Second, our implementation memoizes results of automaton intersection. Since the top-level synthesis algorithm ends up intersecting the same pair of automata many times, this kind of memoization is useful for improving the efficiency of our synthesis procedure. 
In addition to these optimizations, our implementation also limits the number of nested $\getcell$ constructs to be at most $4$. 
We have found this restriction to improve the scalability of automaton intersection without affecting expressiveness in practice.

\section{Evaluation}\label{sec:eval}

In this section, we present the results of our evaluation on $84$ data completion benchmarks collected from online forums.
All experiments are conducted on an Intel Xeon(R) machine with an E5-2630 CPU and 64G of memory. 
%All \tool experiments are conducted on a MacBook Pro with Intel Core i7 processor and 16G of memory running on the OS X Yosemite operating system. 

\vspace{0.05in}\noindent
{\bf \emph{Benchmarks.}}
To evaluate the proposed synthesis technique , we collected a total of $84$ data completion benchmarks from StackOverflow by searching for posts containing relevant keywords such as \emph{``data imputation", ``missing value", ``missing data", ``spreadsheet formula"}, and so on. 
We then manually inspected each post and only retained those benchmarks that are indeed relevant to data completion and that contain at least one example. 
Among these $84$ benchmarks, $46$ involve data imputation in languages such as R and Python, $32$ perform spreadsheet computation in Excel and Google Sheets, and $6$ involve data completion in relational databases.

Recall that an input to \tool consists of (a) a small example table, (b) a sketch formula, and (c) a mapping from each hole in the sketch to a set of examples of the form $i \mapsto [o_1, \ldots, o_n]$. As it turns out, most posts  contain exactly the type of information that \tool requires: Most questions related to data completion already come with a small example table, a simple formula (or a short description in natural language), and a few examples that show how to instantiate the formula for concrete cells in the table. 

We categorize our benchmarks in 21 groups according to their shared functionality. Specifically, as shown in ~\figref{benchmarks}, benchmarks in the same category typically require the same sketch. For instance, all benchmarks in  the first category in~\figref{benchmarks} share the sketch $\sketchhole_1$ and require filling the missing value with the previous/next non-missing value (with or without the same key). %For instance, both \exref{locf} and \exref{locfid} from Section~\ref{sec:examples} belong to this category. 
The column labeled \emph{``Formula sketch"} shows the concrete sketch for each benchmark category. Observe that most of these sketches are extremely simple to write. In fact, for over 50\% of the benchmarks, the user only needs to specify the sketch $\sketchhole_1$, which is equivalent to having no sketch at all. 
The next column labeled \emph{Count} indicates the number of benchmarks that belong to the corresponding category, and the column  called \emph{``Avg. table size''} shows the average number of cells in the tables for each  category. 
Finally, the last column labeled \emph{``Avg. \# examples per hole''} shows the average number of examples that the user provides in the original StackOverflow post. Observe that this is \emph{not} the number of examples that \tool actually requires to successfully perform synthesis, but rather the number of all available examples in the  forum post.

%Among all the benchmarks, $14$ require to use conditional programs, $19$ need to use $\filterconstruct$ programs, $18$ involve nested cell programs, $16$ require having predicates that use concrete values in the table, and $25$ involve using relational predicates that check values in two cells. 

\newcounter{bcounter}
\newcommand{\counter}{\stepcounter{bcounter}\thebcounter}
\newcommand{\anewline}{\hline}

\begin{figure*}[t]
\footnotesize
\centering
\begin{tabular}{ || c m{9cm} | c | c | c | c || }
\hline
\multicolumn{2}{ || >{\centering\arraybackslash} m{8cm} | }{\hspace{70pt} Benchmark category description} & Formula sketch & \rot{Count} & \rot{\parbox{2cm}{\centering Avg. table size}} & \rot{\parbox{2cm}{\centering Avg. \# examples per hole}} \\ \hline\hline
\counter & Fill missing value by previous/next non-missing value with/without same keys. & $\sketchhole_1$ & 24 & 24.4 & 5.3 \\ \anewline
\counter & Fill missing value by previous (next) non-missing value with/without same keys if one exists, otherwise use next (previous) non-missing value & $\sketchhole_1$ & 9 & 25.6 & 5.7 \\ \anewline
\counter & Replace missing value by the average of previous and next non-missing values. & $\texttt{AVG}(\sketchhole_1, \sketchhole_2)$ & 3 & 12.7 & 2.3 \\ \anewline
\counter & Fill missing value by the average of previous and next non-missing values, but if either one does not exist, fill by the other one. & $\texttt{AVG}(\sketchhole_1)$ & 2 & 21.5 & 4 \\ \anewline 
\counter & Replace missing value by the sum of previous non-missing value (with or without the same key) and a constant. & $\texttt{SUM}(\sketchhole_1, c)$ & 3 & 31.3 & 5.7 \\ \anewline 
\counter & Replace missing value by the average of all non-missing values in the same row/column (with or without same keys). & $\texttt{AVG}(\sketchhole_1)$ & 7 & 21.7 & 3.1 \\ \anewline
\counter & Replace missing value by the max/min of all non-missing values in the column with the same key. & $\texttt{MAX}(\sketchhole_1), \  \texttt{MIN}(\sketchhole_1)$ & 2 & 28.0 & 3 \\ \anewline
\counter & Fill missing value by  linear interpolation of previous/next non-missing values. & $\texttt{INTERPOLATE}(\sketchhole_1, \sketchhole_2)$ & 2 & 28.0 & 7.5 \\ \anewline 
\counter & Fill cells by copying values from other cells in various non-trivial ways, such as by copying the first/last entered entry in the same/previous/next row, and etc. & $\sketchhole_1$ & 13 & 44.5 & 10.2 \\ \anewline
\counter & Fill value by the sum of a range of cells in various ways, such as by summing all values to the left with the same keys. & $\texttt{SUM}(\sketchhole_1)$ & 4 & 47.8 & 10.3 \\ \anewline 
\counter & Fill cells with the count of non-empty cells in a range. & $\texttt{COUNT}(\sketchhole_1)$ & 1 & 32.0 & 3 \\ \anewline 
\counter & Fill cells in a column by the sum of values from two other cells. & $\texttt{SUM}(\sketchhole_1, \sketchhole_2)$ & 2 & 38.3 & 6.5 \\ \anewline 
\counter & Fill each value in a column by the difference of values in two other cells in different columns found in various ways. & $\texttt{MINUS}(\sketchhole_1, \sketchhole_2)$ & 4 & 39.0 & 3.5 \\ \anewline 
\counter & Replace missing value by the average of two non-missing values to the left. & $\texttt{AVG}(\sketchhole_1, \sketchhole_2)$ & 1 & 32.0 & 5 \\ \anewline 
\counter & Complete a column so that each value is the difference of the sum of a range of cells and another fixed cell. & $\texttt{MINUS}(\texttt{SUM}(\sketchhole_1), \sketchhole_2)$ & 1 & 27.0 & 8 \\ \anewline 
\counter & Fill each value in a column by the difference of a cell and sum of a range of cells. & $\texttt{MINUS}(\sketchhole_1, \texttt{SUM}(\sketchhole_2))$ & 1 & 10.0 & 3 \\ \anewline
\counter & Create column where each value is the max of previous five cells in  sibling column. & $\texttt{MAX}(\sketchhole_1)$ & 1 & 60.0 & 15 \\ \anewline 
\counter & Fill blank cell in a column by concatenating two values to its right. & $\texttt{CONCAT}(\sketchhole_1, \sketchhole_2)$ & 1 & 12.0 & 2 \\ \anewline 
\counter & Fill missing value by the linear extrapolation of the next two non-missing values to the right, but if there is only one or zero such entries, fill by the linear extrapolation of the previous two non-missing values to the left. & $\texttt{EXTRAPOLATE}(\sketchhole_1)$ & 1 & 121.0 & 16 \\ \anewline
\counter & Replace missing values by applying an equation (provided by the user) to the previous and next non-missing values. & $\texttt{SUM}(\sketchhole_1, \frac{\texttt{MINUS}(\sketchhole_1, \sketchhole_2)}{\texttt{ROW}(\sketchhole_2) - \texttt{ROW}(\sketchhole_1)})$ & 1 & 60.0 & 9 \\ \anewline 
\counter & Fill missing value using the highest value or linear interpolation of two values before and after it, based on two different criteria. & --- & 1 & 60.0 & 10 \\ \anewline 
\hline
\multicolumn{3}{|| c |}{Summary} & 84 & 32.0 & 6.3 \\ \hline
\end{tabular}
\caption{Benchmark statistics.}
\figlabel{benchmarks}
\vspace{-0.1in}
\end{figure*}

\vspace{0.05in}\noindent
{\bf \emph{Experimental Setup.}}
Since \tool is meant to be used in an interactive mode where the user iteratively provides more examples, we simulated a realistic usage scenario in the following way: First, for each benchmark, we collected the set $\mathcal{S}$ of all examples provided by the user in the original Stackoverflow post. We then randomly picked a single example $e$ from $\mathcal{S}$ and used \tool to synthesize a program $P$ satisfying $e$. If $P$ failed any of the examples in $\mathcal{S}$, we then randomly sampled a failing test case $e'$ from $\mathcal{S}$ and used \tool to synthesize a program that satisfies both $e$ and $e'$. We repeated this process of randomly sampling examples from $\mathcal{S}$ until either (a) the synthesized program $P$ satisfies all examples in $\mathcal{S}$, or (b) we exhaust all examples in $\mathcal{S}$, or (c) we reach a time-out of $30$ seconds per synthesis task. At the end of this process, we manually inspected the program $P$ synthesized by \tool and checked whether $P$ conforms to the natural language description provided by the user.

\newcolumntype{?}{!{\vrule width 1.1pt}}

\setcounter{bcounter}{0}

\begin{figure*}[t]
\footnotesize
\centering
\begin{tabular}{ || c | c || c | c  c | c  c ? c | c  c | c  c ? c | c  c | c | c  c || }
\hline
\multirow{8}{*}[1pt]{\rott{90}{Benchmark category}} 
& \multirow{8}{*}{\rott{90}{Count}} 
& \multicolumn{5}{ c ?}{\textbf{\tool}} 
& \multicolumn{5}{ c ?}{\textbf{\prose}} 
& \multicolumn{6}{ c ||}{\textbf{\sketchTool}} \\ \cline{3-18}
& & \multirow{2}{*}{\rott{90}{\# Solved}} 
& \multicolumn{2}{c |}{\rott{65}{\parbox{1.8cm}{\centering Running time per benchmark (sec)}}} 
& \multicolumn{2}{c ?}{\rott{65}{\parbox{1.8cm}{\centering \# Examples used per hole}}} 
& \multirow{2}{*}{\rott{90}{\# Solved}} 
& \multicolumn{2}{c |}{\rott{65}{\parbox{1.8cm}{\centering Running time per benchmark (sec)}}} 
& \multicolumn{2}{c ?}{\rott{65}{\parbox{1.8cm}{\centering \# Examples used per hole}}}
& \multirow{2}{*}[4ex]{\rott{90}{\# Solved (2 exs)}}
& \multicolumn{2}{c |}{\rott{65}{\parbox{1.8cm}{\centering Running time per benchmark (sec)}}}
& \multirow{2}{*}[4ex]{\centering\rott{90}{\# Solved (3 exs)}}
& \multicolumn{2}{c ||}{\rott{65}{\parbox{1.8cm}{\centering Running time per benchmark (sec)}}}
\\ \cline{4-7} \cline{9-12} \cline{14-15} \cline{17-18}
& & & Avg. & Med. & Avg. & Med. & & Avg. & Med. & Avg. & Med. & & Avg. & Med. & & Avg. & Med. \\ 
\hline\hline 
\counter & 24 & 24 & 0.41 & 0.04 & 1.1 & 1.0 & 24 & 1.32 & 0.73 & 1.1 & 1.0 & 6 & 230 & 224 & 6 & 314 & 281 \\ \anewline
\counter & 9 & 9 & 0.50 & 0.13 & 2.7 & 3.0 & 7 & 4.88 & 1.13 & 2.4 & 2.0 & 2 & 182 & 182 & 0 & --- & --- \\ \anewline
\counter & 3 & 3 & 0.05 & 0.04 & 1.0 & 1.0 & 3 & 5.16 & 5.89 & 1.0 & 1.0 & 0 & --- & --- & 0 & --- & --- \\ \anewline
\counter & 2 & 2 & 0.19 & 0.19 & 2.0 & 2.0 & 1 & 2.11 & 2.11 & 2.0 & 2.0 & 0 & --- & --- & 0 & --- & --- \\ \anewline 
\counter & 3 & 3 & 0.18 & 0.14 & 1.3 & 1.0 & 3 & 0.90 & 0.99 & 1.7 & 1.0 & 0 & --- & --- & 0 & --- & ---\\ \anewline 
\counter & 7 & 6 & 0.09 & 0.07 & 1.8 & 2.0 & 5 & 15.86 & 8.31 & 1.8 & 2.0 & 5 & 353 & 352  & 4 & 399 & 400 \\ \anewline
\counter & 2 & 2 & 0.66 & 0.66 & 2.0 & 2.0 & 1 & 296.17 & 296.17 & 3.0 & 3.0 & 0 & --- & --- & 0 & --- & --- \\ \anewline
\counter & 2 & 2 & 0.15 & 0.15 & 1.0 & 1.0 & 1 & 19.72 & 19.72 & 1.0 & 1.0 & 1 & 501 & 501 & 0 & --- & --- \\ \anewline 
\counter & 13 & 10 & 1.55 & 0.31 & 2.8 & 2.0 & 5 & 6.02 & 1.52 & 1.4 & 1.0 & 2 & 507 & 507 & 0 & --- & --- \\ \anewline
\counter & 4 & 3 & 0.42 & 0.30 & 1.7 & 2.0 & 1 & 2.27 & 2.27 & 2.0 & 2.0 & 0 & --- & --- & 0 & --- & --- \\ \anewline 
\counter & 1 & 1 & 0.59 & 0.59 & 1.0 & 1.0 & 0 & --- & --- & --- & --- & 3 & 223 & 182 & 3 & 353 & 298 \\ \anewline 
\counter & 2 & 2 & 0.51 & 0.51 & 1.0 & 1.0 & 1 & 66.95 & 66.95 & 2.0 & 2.0 & 0 & --- & --- & 0 & --- & --- \\ \anewline 
\counter & 4 & 4 & 0.51 & 0.46 & 2.0 & 2.0 & 2 & 1.52 & 1.52 & 2.0 & 2.0 & 0 & --- & --- & 0 & --- & --- \\ \anewline 
\counter & 1 & 1 & 0.16 & 0.16 & 3.0 & 3.0 & 0 & --- & --- & --- & --- & 0 & --- & --- & 0 & --- & --- \\ \anewline 
\counter & 1 & 1 & 0.11 & 0.11 & 2.0 & 2.0 & 1 & 148.95 & 148.95 & 3.0 & 3.0 & 0 & --- & --- & 0 & --- & --- \\ \anewline 
\counter & 1 & 1 & 0.03 & 0.03 & 2.0 & 2.0 & 0 & --- & --- & --- & --- & 0 & --- & --- & 0 & --- & --- \\ \anewline
\counter & 1 & 1 & 1.96 & 1.96 & 4.0 & 4.0 & 1 & 183.19 & 183.19 & 2.0 & 2.0 & 1 & 78 & 78 & 1 & 81 & 81 \\ \anewline 
\counter & 1 & 1 & 0.01 & 0.01 & 1.0 & 1.0 & 1 & 1.44 & 1.44 & 1.0 & 1.0 & 0 & --- & --- & 0 & --- & --- \\ \anewline 
\counter & 1 & 1 & 13.66 & 13.66 & 5.0 & 5.0 & 0 & --- & --- & --- & --- & 0 & --- & --- & 0 & --- & ---\\ \anewline
\counter & 1 & 1 & 1.92 & 1.92 & 1.0 & 1.0 & 0 & --- & --- & --- & --- & 0 & --- & --- & 0 & --- & --- \\ \anewline 
\counter & 1 & 0 & --- & --- & --- & --- & 0 & --- & --- & --- & --- & 0 & --- & --- & 0 & --- & --- \\ \anewline 
\hline
All & 84 & 78 & 0.70 & 0.19 & 1.8 & 2.0 & 57 & 16.09 & 1.18 & 1.5 & 1.0 & 20 & 289 & 226 & 14 & 330 & 314 \\ \hline
\end{tabular}
\vspace{-0.1in}
\caption{Evaluation results of \tool, as well as $\prose$ and $\textsc{SKETCH}$ as two baselines. The time-out for \tool is set to be $30$ seconds, whereas the time-out for baselines is set to be $10$ minutes.}
\figlabel{expresults}
\vspace{-0.1in}
\end{figure*}

\vspace{0.05in}\noindent
{\bf \emph{Results.}}
We present the main results of our evaluation of \tool in \figref{expresults}. 
The column \emph{``\# Solved''} shows the number of benchmarks that can be successfully solved by \tool for each benchmark category. 
Overall, \tool can successfully synthesize over 92\% of the benchmarks. Among the six benchmarks that cannot be synthesized by \tool, one benchmark (Category 21) cannot be expressed using our specification language. For the remaining 5 benchmarks, \tool fails to synthesize the correct program due to limitations of our DSL, mainly caused by the restricted vocabulary of predicates. For instance, two benchmarks require capturing the concept ``nearest", which is not expressible by our current predicate language.

Next, let us consider the running time of \tool, which is shown in the column labeled \emph{``Running time per benchmark"}. We see that \tool is quite fast in general and takes an average of $0.7$ seconds to solve a benchmark. The median time to solve these benchmarks is $0.19$ seconds. In cases where the sketch contains multiple holes,  the reported running times include the time to synthesize \emph{all} holes in the sketch. In more detail, \tool can synthesize 75\% of the benchmarks in under one second and 87\% of the benchmarks in under three seconds. There is one benchmark (Category 19) where \tool's running time exceeds $10$ seconds. This is because (a) the size of the example table provided by the user is  large in comparison to other example tables, and (b) the table contains over 100 irrelevant strings that form the universe of constants used in predicates. These irrelevant entries cause \tool to consider over $30,000$ predicates to be used in the $\getcell$ and $\filterconstruct$ programs.

Finally, let us look at the number of examples used by \tool, as shown in the column labeled \emph{``\# Examples used per hole''}. As we can see, the number of examples used by \tool is much smaller than the total number of examples provided in the benchmark (as shown in~\figref{benchmarks}). Specifically, while StackOverflow users provide about $6$ examples on average, \tool requires only about $2$ examples to synthesize the correct program. This statistic highlights that \tool can effectively learn general programs from very few input-output examples.

%The average time for FTA construction is $0.48$ seconds per benchmark, and the average time for synthesis is $0.21$ seconds per benchmark. 

\vspace{0.05in}\noindent
{\bf \emph{Comparison with $\prose$.}} In this paper, we argued that our proposed FTA-based technique can be viewed as a new version space learning algorithm; hence, we also empirically compare our approach again \prose~\cite{flashmeta}, which is the state-of-the-art version space learning framework that has been deployed in Microsoft products. To provide some background, \prose propagates example-based constraints on subexpressions using the inverse semantics of DSL operators and then represents all programs that are consistent with the examples using a version space algebra (VSA) data structure~\cite{vsa}.

To allow a fair comparison between \prose and \tool, we use the same algorithm presented in Section~\ref{sec:synthesis} to learn the $\seqconstruct$ construct (i.e., branches), but we encode simple programs in the DSL using the \prose format.~\footnote{ \prose performs significantly worse (i.e., terminates on only three benchmarks) if we use \prose's built-in technique for learning $\seqconstruct$.} Since \prose's learning algorithm requires so-called \emph{witness functions}, which describe the inverse semantics of each DSL construct, we also manually wrote precise witness functions for all constructs in our DSL.  Finally, we use the same scoring function $\theta$ described in the Appendix to rank different programs in the version space. 

The results of our  evaluation  are presented under the \prose column in \figref{expresults}. Overall, \prose can successfully solve 68\% of the benchmarks in an average of 15 seconds, whereas \tool can solve 92\% of the benchmarks in an average of 0.7 seconds. These results indicate that \tool is superior to \prose, both in terms of its running time and the number of benchmarks that it can solve. Upon further inspection of the \prose results, we found that the tasks that can be automated using \prose tend to  be relatively simple ones, where  the input table size is very small or the desired program is relatively simple. For benchmarks that have larger tables or involve more complex synthesis tasks (e.g., require the use of $\filterconstruct$ operator), \prose does not scale well -- i.e., it might take much longer time than \tool, time out in 10 minutes, or run out of memory.  We provide more details and intuition regarding why our FTA-based learning algorithm performs better than \prose's VSA-based algorithm in Section~\ref{sec:vsa}.

The careful reader may have observed in \figref{expresults} that \prose requires fewer examples on average than \tool (1.5 vs. 1.8). However, this number is quite misleading, as the benchmarks that can be solved using \prose are relatively simple and therefore require fewer examples on average.

\vspace{0.05in}\noindent
{\bf \emph{Comparison with $\sketchTool$.}} Since our synthesis methodology involves a sketching component in addition to examples, we also compare \tool against $\sketchTool$, which is the state-of-the-art tool for program sketching. To compare \tool against \sketchTool, we define the DSL operators using nested and recursive structures in $\sketchTool$. For each struct, we define two corresponding functions, namely \t{RunOp} and \t{LearnOp}. The \t{RunOp} function defines the semantics of the operator whereas  \t{LearnOp} encodes a \sketchTool generator that defines the bounded space of all possible  expressions in the DSL. The specification is encoded as a sequence of assert statements of the form \t{assert} RunExtractor(LearnExtractor(), $i_k$) == $L_k$, where ($i_k, L_k$) denotes the input-output examples. To optimize the sketch encoding further, we use the input-output examples inside the \t{LearnOp} functions, and we also manually unroll and limit the recursion in predicates and cell programs to 3 and 4 respectively.

When we use the complete DSL encoding,  \sketchTool  was able to solve only 1 benchmark out of 84 within a time limit of 10 minutes per benchmark. We then simplified the \sketchTool encoding by removing the \t{Seq} operator, which allows us to synthesize only conditional-free programs.  As shown in~\figref{expresults}, \sketchTool  terminated on 20 benchmarks within  10 minutes using 2 input-output examples. The average time to solve each benchmark was 289 seconds. However, on manual inspection, we found that most of the synthesized programs were not the desired ones. When we increase the number of input-output examples to 3, 14 benchmarks terminated with an average of 330 seconds, but only 5 of these 14 programs were the desired ones. 
We believe that \sketchTool performs poorly due to two reasons: First, the constraint-based encoding in \sketchTool does not scale for complex synthesis tasks that arise in the data completion domain. 
Second, since it is difficult to encode our domain-specific ranking heuristics using primitive cost operations supported by \sketchTool, it often generates undesired programs. 
\begin{comment}
\xinyu{Second, since it is not possible to encode domain-specific ranking heuristics in \sketchTool, it often generates the incorrect program.}
\end{comment}
In summary, this experiment confirms that a general-purpose program sketching tool is not adequate for automating the kinds of data completion tasks that arise in practice.

\section{Version Space Learning using Finite Tree Automata}\label{sec:vsa}
So far in this paper, we focused on our algorithm for synthesizing data completion tasks in our domain-specific language. However, as argued in Section~\ref{sec:intro}, our FTA-based formulation of unification can be seen as a new version-space learning algorithm. In this section, we outline how our FTA-based learning algorithm could be applied to other settings, and we also discuss the advantages of our learning algorithm compared to prior VSA-based techniques~\cite{vsa, flashfill, flashmeta}.

%In this section, we present our general idea of performing version space learning using finite tree automata (FTA). 
%We also compare our idea against prior version space learning techniques, such as techniques that use Version Space Algebra~\cite{vsa, flashfill, flashmeta} for program synthesis. 

\subsection{The general idea}\label{sec:general-fta}

To see how our FTA-based unification procedure can be used as a general version space learning algorithm, let us consider a  domain-specific language specified by a context-free grammar $G = (T, N, P, S)$, where $T$ is a finite set of terminals (i.e., variables and constants), $N$ is the set of non-terminal symbols, $P$ is a set of productions of the form $e \rightarrow F(e_1, \ldots, e_n)$ where $F$ is a built-in DSL function (i.e., ``component"),   and $S$ is the start symbol representing a complete program.   To simplify the presentation, let us assume that the components used in each production are first-order; if they are higher-order, we can combine our proposed methodology with enumerative search (as we did in this paper for dealing with predicates inside the $\filterconstruct$ and $\getcell$ constructs).

Now, our general version space learning algorithm works as follows. For each input-output example  $\sigma \rightarrow o$, where $\sigma$ is a valuation and $o$ is the output value, we construct an FTA $\automaton = (\states, \alphabet, \finalstates, \transitionrules)$ that represents exactly the set of programs that are consistent with the examples. Here, the alphabet $\alphabet$ of the FTA consists of the built-in components provided by the DSL. 

To construct the states $\states$ of the FTA, let us assume that every non-terminal symbol $n \in N$ has a pre-defined universe $U_n = \{v_1, \ldots, v_k\}$ of values that it can take. Then, we introduce a state $q_{n}^{v_i}$ for every $n \in N$ and $v_i \in U_n$; let us refer to the set of states for all non-terminals in $N$ as $Q_N$. We also construct a set of states $Q_T$ by introducing a state $q_t$ for each terminal $t \in T$. Then, the set of states in the FTA is given by $Q_N \cup Q_T$.

Next, we construct the transition rules $\transitionrules$ using the productions $P$ in the grammar. To define the transitions, let us define a function $dom(s)$ that gives the domain of $s$ for every symbol $s$:
\[
dom(s) = \left \{
\begin{array}{ll}
s & {\rm if} \ s \ {\rm is \  constant} \\
\sigma(s) & {\rm if} \ s \ {\rm is \ a \ variable} \\
U_s & {\rm if} \ s \ {\rm is \ a \ non\text{-}terminal, and \ } U_s {\rm \ is \ its \ corresponding \ universe} \\
\end{array}
\right .
\]
Now, consider a production of the form $n \rightarrow F(s_1, \ldots, s_k)$ in the grammar where $n$ is a non-terminal and each $s_i$ is either a terminal or non-terminal. For every $v_i \in dom(s_i)$, we add a transition $F(q_{s_1}^{v_1}, \ldots, q_{s_k}^{v_k}) \rightarrow q_{n}^v$ iff we have $v = \semantics{F(v_1, \dots, v_k)}$. In addition, for every variable $x$, we add a transition $x \rightarrow q_x$. Finally, the final state $\finalstates$ is a singleton containing the state $q_S^o$, where $S$ is the start symbol of the grammar and $o$ is the output in the example. 

Given this general methodology for FTA construction, the learning algorithm works by constructing the FTA for each individual example and then intersecting them. The final FTA represents the version space of all programs that are consistent with the examples.

\subsection{Comparison with prior version space learning techniques}

As mentioned briefly in Section~\ref{sec:intro}, we believe that our FTA-based learning algorithm has two important advantages compared to the VSA-based approach in \prose. First, FTAs yield a more succinct representation of the version space compared to VSAs in \prose. To see why, recall that VSA-based approaches construct more complex version spaces by combining smaller version spaces using algebraic operators, such as Join and Union. In essence, \prose constructs a hierarchy of version spaces where the version spaces at lower levels can be shared by version spaces at higher levels, 
but cyclic dependencies between version spaces are not allowed. As a result, \prose must unroll recursive language constructs to introduce new version spaces, but this unrolling leads to a less compact representation with less sharing between version spaces at different layers. The second advantage of our learning algorithm using FTAs is that it does not require complex \emph{witness functions} that encode inverse semantics of DSL constructs. Specifically, since \prose propagates examples backwards starting from the output, the developer of the synthesizer must manually specify witness functions. In contrast, the methodology we outlined in Section~\ref{sec:general-fta} does not require any additional information beyond the grammar and semantics of the DSL.

We refer the interested reader to the Appendix for an example illustrating the differences between \prose's VSA-based learning algorithm and our FTA-based technique.

\section{Related work}
In this section, we compare and contrast our approach with prior work on program synthesis.

\vspace{0.1in}\noindent
{\bf \emph{Programming-by-example.}}
In recent years, there has been significant interest in programming by example (PBE)~\cite{flashfill, flashmeta, fidex, lambda2, mapreduce, myth, metasketch, transit, synquid,hades}.
Existing PBE techniques can be roughly categorized into two classes, namely \emph{enumerative search} techniques~\cite{lambda2, myth, transit}, and those based on \emph{version space learning}~\cite{flashfill, flashmeta, number}.

The enumerative techniques search a space of programs to find a \emph{single} program that is consistent with the examples. Specifically, they enumerate all programs in the language in a certain order and terminate when they find a program that satisfies all examples. Recent techniques in this category employ various pruning methods and heuristics, for instance by using type information~\cite{synquid, myth}, checking partial program equivalence~\cite{transit, aws13cav},  employing deduction~\cite{lambda2}, or performing stochastic search~\cite{stoke}. 

In contrast, PBE techniques based on version space learning construct a compact data structure representing \emph{all} possible programs that are consistent with the examples. 
%~\cite{flashfill,flashrelate,semanticstring,flashextract,fidex,number,datatype,flashmeta}. 
%The programs that are consistent with \emph{all} examples are then obtained by taking the intersection of the individual data structures. 
The notion of \emph{version space} was originally introduced by Mitchell~\cite{vs} as a general search technique for learning boolean functions from samples.
%and a language over which to search.
%The basic idea is to maintain a set of all boolean functions that could be the desired unknown function (referred to as version space). 
Lau et al. later extended this concept to \emph{version space algebra} for learning more complex functions~\cite{vsa}. The basic idea is to build up a complex version space by \emph{composing} together version spaces containing simpler functions, thereby representing hypotheses  hierarchically~\cite{robot}.  

%Recent papers on PBE also adopt this methodology to synthesize programs for string transformations~\cite{flashfill, semanticstring}, data extraction~\cite{flashextract}, data filtering~\cite{fidex}, number transformations~\cite{number}, and data type transformations~\cite{datatype}. VSA has also been used for generating domain-specific synthesizers~\cite{flashmeta}. 

The synthesis algorithm proposed in this paper is another technique for performing version space learning -- i.e., we build a data structure (namely, finite tree automaton) that represents all consistent hypotheses. However, our approach differs from previous work using version space learning in several key aspects:
First, unlike VSA-based techniques that decompose the version space of complex programs into smaller version spaces of simpler programs, we directly construct a tree automaton whose language accepts all consistent programs. 
Second, our FTA construction is done in a forward manner, rather than by back-propagation as in previous work~\cite{flashmeta}. 
Consequently, we believe that our technique results in a more compact representation and enables better automation.

\vspace{0.1in}\noindent
{\bf \emph{Program sketching.}}
In sketch-based synthesis~\cite{sketching, solarthesis, comsketching, stencils}, the programmer provides a skeleton of the program with missing expressions (holes).  Our approach is similar to program sketching in that we require the user to provide a formula sketch, such as ${\tt SUM}(\sketchhole_1, \sketchhole_2)$. However, holes in our formula sketches are  programs rather than constants. Furthermore, while $\sketchTool$ uses a constraint-based counter-example guided inductive synthesis algorithm, $\tool$ uses a combination of finite tree automata and enumerative search. As we show in our experimental evaluation, $\tool$ is significantly more efficient at learning data completion programs  compared to $\sketchTool$.

\vspace{0.1in}\noindent
{\bf \emph{Tree automata.}}
Tree automata were introduced in the late sixties as a generalization of finite word automata~\cite{treeautomata}. Originally, ranked tree automata were used to prove the existence of a decision procedure for weak monadic second-order logic with multiple successors~\cite{treeautomata}. In the early 2000s, unranked tree automata have also gained popularity as a tool for analyzing XML documents where the number of children of a node are not known a priori~\cite{unranked1,unranked2}. More recently, tree automata have found numerous applications in the context of software verification~\cite{treeverification,treeverification1}, analysis of XML documents~\cite{treexml,unranked1,unranked2}, and natural language processing~\cite{treenlp,treenlp1}. 
Most related to our technique is the work of  Parthasarathy  in which they advocate the use of tree automata as a theoretical basis for synthesizing reactive programs~\cite{madhu11csl}. In that work, the user provides a regular $\omega$-specification describing the desired reactive system, and the proposed synthesis methodology constructs a non-deterministic tree automaton representing programs (over a simple imperative language) that meet the user-provided specification. The technique first constructs an automaton that accepts reactive programs corresponding to the negation of the regular $\omega$-specification, and then complements it to obtain the automaton for representing the desired set of programs. In contrast to the purely theoretical work of Parthasarathy in the context of synthesizing reactive programs from regular $\omega$-specifications, we show how finite tree automata can be used in the context of program synthesis from examples. Moreover, we combine this FTA-based approach with enumerative search to automatically synthesize programs for real-world data completion tasks in a functional DSL with higher-order combinators.

\section{Conclusions and Future Work}
In this paper, we presented a new approach for automating data completion tasks using a combination of \emph{program sketching} and \emph{programming-by-example}. Given a formula sketch where holes represent programs and a set of input-output examples for each hole, our technique generates a script that can be used to automate the target data completion task. To solve this problem, we introduced a new domain-specific language that combines relational and spatial reasoning for tabular data and a new synthesis algorithm for generating programs over this DSL. Our synthesis procedure combines enumerative search (for learning conditionals) with a new version-space learning algorithm that uses finite tree automata. We also showed the generality of our FTA-based learning algorithm  by explaining how it can be used synthesize programs over any arbitrary DSL specified using a context-free grammar. 

We evaluated our proposed synthesis algorithm on 84 data completion tasks collected from StackOverflow and compared our approach with two existing state-of-the-art synthesis tools, namely \prose and \sketchTool. Our experiments  demonstrate that  \tool is practical enough to automate a large class of data completion tasks and that it significantly outperforms both \prose and \sketchTool in terms of both running time and the number of benchmarks that can be solved.

We are interested in two main directions for future work. First, as discussed in Section~\ref{sec:eval}, there are a few benchmarks for which \tool's DSL (specifically, predicate language) is not sufficiently expressive. While such benchmarks seem to be  relatively rare, we would like to investigate how to enrich the DSL so that all of these tasks can be automated. Second, we would like to apply our new version-learning algorithm using FTAs to other domains beyond data completion. We believe that our new VS-learning algorithm can be quite effective in other domains, such as automating table transformation tasks.

%would like to investigate the applicability of finite tree automata in domains othe

\begin{comment}
We plan to apply the techniques proposed in this paper to more scenarios other than data completion. 
One current limitation of \tool is that the predicate language only contains conjunctive string equalities (up to size $3$). 
Although this is not a serious problem in the domain of data completion, we envision it would be necessary to improve the predicate language expressiveness by extending its vocabulary to support more features, such as substring matching. 
As a result, this might lead to potential scalability issues, which we plan to address by designing predicate learning algorithms or using symbolic encoding. 
Another idea we plan to work on is to improve the top-level synthesis algorithm, since the worst-case running time is $\mathcal{O}(n^m)$ in the number of examples $n$ and the number of branches $m$ in the learned program. In the domain of data completion, this is not a problem due to the effectiveness of a ranking algorithm. However, for situations where it is hard to predicate a correct program with few examples, the efficiency of the synthesis algorithm would be critical for improving the scalability of the whole system. 
\end{comment}

\newpage

% Bibliography
%\bibliography{references}
\bibliography{main}

\newpage

\section*{Appendix A: Heuristic Scoring Function}

\setcounter{theorem}{0}

Recall that our synthesis algorithm uses a scoring function $\score$ to choose between multiple programs that satisfy the input-output examples.  The design of the scoring fuction follows the \emph{Occam's razor} principle and tries to favor simpler, more general programs over complex ones.

In more detail, our scoring function assign scores to constants, cell mappers, and predicates in our DSL in a way that satisfies the following properties: 
\begin{itemize}
\item 
A predicate with mapper $\lambda c. c$ has a higher score than the same predicate with other mappers. 
\item 
For predicates containing the same mapper $\mapper$, we require that the scoring function satisfies the following constraint:
\[ \score(\true) > \score(\valeqconst{\mapper(z)}{\missingval}) > \score(\valeq{\mapper(y)}{\mapper(z)}) > \score(\valeqconst{\mapper(z)}{s}) = \score(\valneqconst{\mapper(z)}{s}) \]
\item 
$\score(\pred_1, \dots, \pred_n)$ takes into account  both the scores of each conjunct as well as the number of conjuncts. That is, $\score$ assigns a higher score to predicates that have conjuncts with higher scores, and assigns lower scores to predicates with more conjuncts. One design choice satisfying this criterion is to take the average of scores of all the terms in the conjunct. 
\item 
For scores of integer $k$ we have $\score(1) > \score(2) > \score(-1) > \score(3) > \score(-2) > \score(-3)$. 
\end{itemize}

Using the scores assigned to predicates, mappers, and constants, we then assign scores to more complex programs in the DSL in the following way.
The score of a $\getcell$ program is defined by taking into account both the scores of its arguments and the recursion depth (number of nested $\getcell$ constructs). One possible way to assign scores to $\getcell$ programs is therefore the following:
\[ 
\begin{aligned}\nonumber
\score(\getcell(\inputcell, \dir, k, \lambda\startcell\lambda\itercell.\phi)) & \assign \score(k) \cdot \score(\phi) \\ 
\score(\getcell(\cellprog, \dir, k, \lambda\startcell\lambda\itercell.\phi)) & \assign \frac{\score(\cellprog) + \score(k) \cdot \score(\phi)}{\textit{depth}(\cellprog)}  \\ 
\textit{depth}(\inputcell) & \assign 0 \\ 
\textit{depth}(\getcell(\cellprog, \dir, k, \lambda\startcell\lambda\itercell.\phi)) & \assign \textit{depth}(\cellprog) + 1 \\  
\end{aligned}
\]

The score of a simple program, i.e., $\unionconstruct(\cellprog_1, \dots, \cellprog_n)$ or $\filterconstruct(\cellprog_1, \cellprog_2, \cellprog_3, \lambda\startcell\lambda\itercell.\phi)$, takes into account of the scores of its arguments and the number of the arguments. Specifically, it assigns scores in the following way:
\[
\begin{aligned}\nonumber
\score(\unionconstruct(\cellprog_1, \dots, \cellprog_n)) & \assign \frac{\score(\cellprog_1) + \dots + \score(\cellprog_n)}{n} \\ 
\score(\filterconstruct(\cellprog_1, \cellprog_2, \cellprog_3, \lambda\startcell\lambda\itercell.\phi)) & \assign \frac{\score(\cellprog_1) + \score(\cellprog_2) + \score(\cellprog_3)}{3} \\ 
\end{aligned}
\]

\section*{Appendix B: Proof of Soundness and Completeness}

\begin{theorem} {\bf (Soundness and Completeness)}
Let $\automaton$ be the finite tree automaton constructed by our technique for example $i \mapsto L$ and table \textnormal{$\inputtable$}. Then, $\automaton$ accepts the tree that represents a simple program $\simpleprog$ iff \textnormal{$\semantics{\simpleprog}_{\inputtable, i} = L$}.
\end{theorem}

\begin{proof}
We first prove soundness -- i.e., if $\automaton$ accepts the tree $t$ that represents a simple program $\simpleprog$, then we have \textnormal{$\semantics{\simpleprog}_{\inputtable, i} = L$}. We show this by inductively proving (call it $P$) that for any program $p$ whose program tree is a sub-tree $t^\prime$ of $t$ and whose height is at most $h$, we have 
\[
\semantics{p}_{\inputtable, i} = \left \{
\begin{aligned}
& L & \langif ~ \text{Root}(t^\prime) = q^* \\ 
& o & \langif ~ \text{Root}(t^\prime) = q_o \\ 
\end{aligned}
\right.
\]
\noindent
The base case for $h=1$ trivially holds, since we have $p = x$, $t= \state_i$, $t^\prime = \state_i$, and $\semantics{p}_{\inputtable, i} = i$.
For the inductive case, we want to prove that $P$ holds for any sub-tree $t^{\prime}$ of height $h+1$. 
Suppose $t^\prime = f_{\vec{\omega}}(t^\prime_1, \dots, t^\prime_m)$ and consider $t^\prime$'s $m$ child-trees $t^\prime_1, \dots, t^\prime_m$. Because a child-tree $t^\prime_j$ is of height at most $h$, $P$ holds for $t^\prime_j$ according to the inductive hypothesis. Furthermore, we have 
\[
\semantics{f(o_1, \dots, o_m, \vec{\omega})} = \left\{
\begin{aligned}
& L & \langif ~ \text{Root}(t^\prime) = q^* \\ 
& o & \langif ~ \text{Root}(t^\prime) = q_o \\ 
\end{aligned}
\right.
\]
\noindent
according to the rules in \figref{transitionrules}. 
Therefore, $P$ also holds for any sub-tree $t^{\prime}$ of height $h+1$, and for any sub-tree of height at most $h+1$ as well due to the inductive hypothesis.  

Now we turn to the proof of completeness -- i.e., if there is a simple program $\simpleprog$ that \textnormal{$\semantics{\simpleprog}_{\inputtable, i} = L$}, then $\automaton$ accepts $\simpleprog$'s program tree $t$. Consider the evaluation of $\simpleprog$ given input cell $i$. In each step in which it evaluates a function of the form $f(o_1, \dots, o_m, \vec{\omega})$ where we have $o_j \in \textsf{Cells}(\inputtable) \cup \{ \bot \}$, there exists a transition in $\automaton$ that goes from states $q_{o_1}, \dots, q_{o_m}$ to a state that represents the evaluation result with an $m$-ary function $f_{\vec{\omega}}$ (according to the rules in \figref{transitionrules}). Therefore, there exists a tree that is accepted by $\automaton$ (according to our construction) and represents simple program $\simpleprog$. 
\end{proof}

\section*{Appendix C: Complexity}

The complexity of our synthesis algorithm depends on the number of examples $n$, the number of branches $k$ in the target program, and the size of the input table (number of cells), $m$. Specifically, the running time complexity of the algorithm is $\complexity{k^n \cdot (m^2)^n}$. To see where this result comes from, observe that the worst-case  complexity of our FTA construction is $\complexity{|\Delta|}$, where $\Delta$ is the set of transitions.  In our case, $|\Delta|$
is bound by $m^2$ because we can have a transition for each pair of cells. Since the number of examples is $n$ and FTA intersection takes quadratic time in the size of each FTA, the time to unify $n$ examples is bound by $\complexity{(m^2)^n}$. Finally, if the learned program has $k$ branches, our algorithm searches for $k^n$ possible partitions. Thus, an upper bound on the run-time complexity is $\complexity{k^n \cdot (m^2)^n}$. However, in practice, since the constructed FTAs are quite sparse, FTA intersection does not result in a quadratic blow-up and remains roughly linear. Hence, in practice, the complexity is closer to $\complexity{k^n \cdot m^2}$. Furthermore, in PBE systems, the user is expected to provide a small number of examples; otherwise, the technique would be too cumbersome for the user. Therefore, in practice, $n$ is expected to be a small number (at most 5 in our experiments). Finally, since the target programs typically do not have a large number of branches,  $k$ is also  expected to be quite small in practice (at most 3 in our experiments).

\section*{Appendix D: In-depth comparison between Version Space Algebras and Finite Tree Automata}

Let us consider how \prose would solve the simple synthesis problem from \exref{exfta}. For simplicity, let us only consider the DSL shown in \figref{simplesyntax}. 
Here, the top-level construct is a cell program $\cellprog$, which is either the input cell $x$, or a $\getcell$ program whose arguments are chosen from a restricted space. Note that $\getcell$ is recursive, and we assume \prose allows at most 3 $\getcell$ programs to be nested together.

\begin{figure}
\begin{tabular}{c c}
\begin{minipage}{0.46\linewidth}
\begin{minipage}{\linewidth}
\small
$\begin{array}{lll}
\langtext{Cell prog.} ~ \cellprog & \assign & \inputcell ~~ | ~~ \getcell(\cellprog, \dir, k, \lambda \startcell. \lambda \itercell. \true) ~; \\ 
\langtext{Direction} ~ \dir & \assign & \updir ~ | ~ \downdir ~ | ~ \leftdir ~ | ~ \rightdir ~; \\ 
\langtext{int} ~ k & \assign & -1 ~ | ~ 1 ~; \\ 
\end{array}$
\caption{A simple DSL.}
\figlabel{simplesyntax}
\vspace{15pt}
\end{minipage}
\begin{minipage}{\linewidth}
\small
$\begin{array}{lll}
\langtext{Cell prog.} ~ \cellprog & \assign & \inputcell ~ | ~ \getcell(\cellprog_1, \dir, k, \lambda \startcell. \lambda \itercell. \true) ~; \\ 
\langtext{Cell prog.} ~ \cellprog_1 & \assign & \inputcell ~ | ~ \getcell(\cellprog_2, \dir, k, \lambda \startcell. \lambda \itercell. \true) ~; \\ 
\langtext{Cell prog.} ~ \cellprog_2 & \assign & \inputcell ~ | ~ \getcell(\inputcell, \dir, k, \lambda \startcell. \lambda \itercell. \true) ~; \\ 
\langtext{Direction} ~ \dir & \assign & \updir ~ | ~ \downdir ~ | ~ \leftdir ~ | ~ \rightdir ~; \\ 
\langtext{int} ~ k & \assign & -1 ~ | ~ 1 ~; \\ 
\end{array}$
\caption{The unrolled grammar in \prose.}
\figlabel{unrolled}
\end{minipage}
\end{minipage}
\begin{minipage}{0.5\linewidth}
\centering
\includegraphics[scale=0.26]{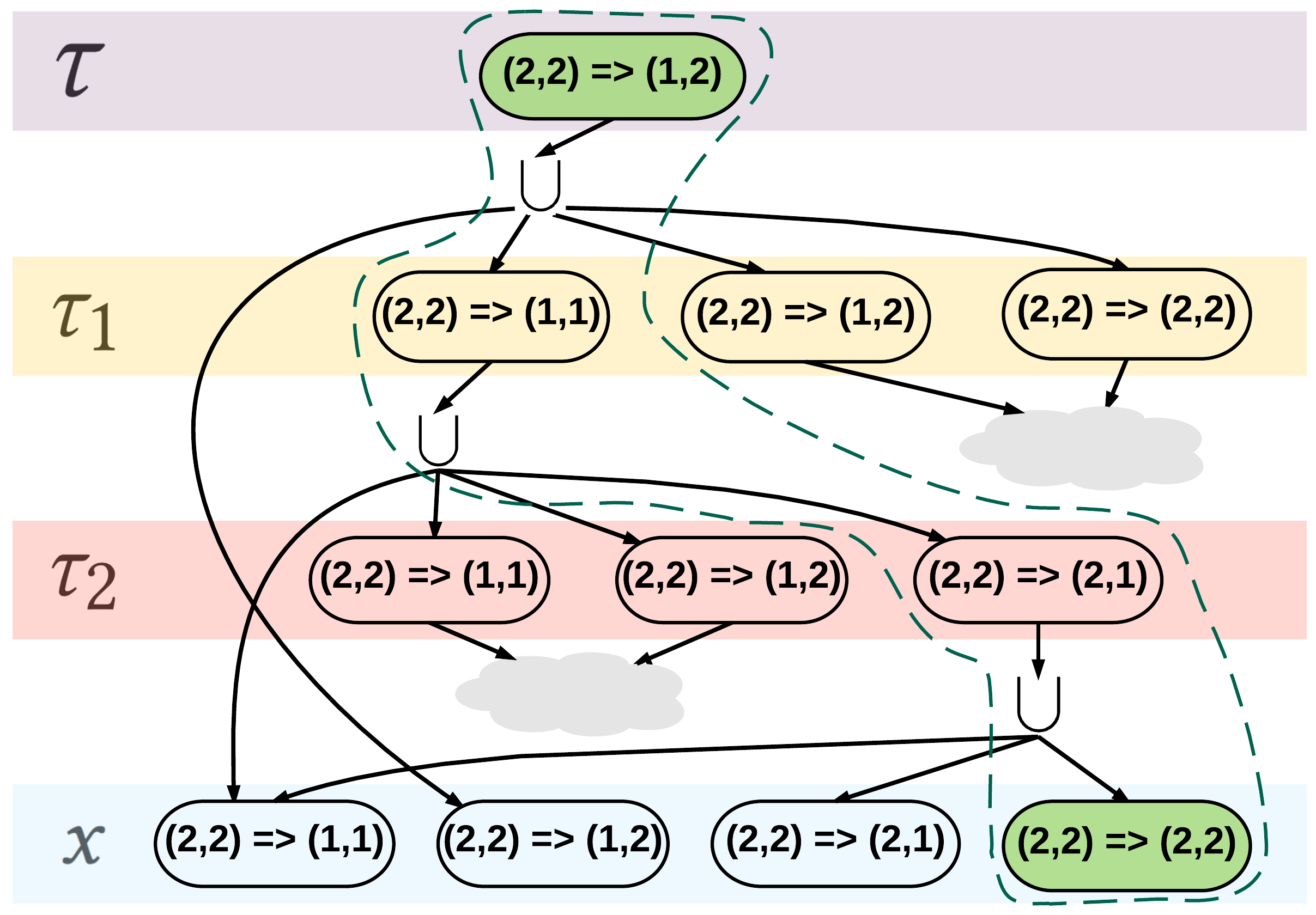}
\caption{Part of the VSA in \prose.}
\figlabel{partvsa}
\end{minipage}
\end{tabular}
\vspace{-10pt}
\end{figure}

\figref{partvsa} shows how \prose constructs the VSA. 
Conceptually, \prose first performs backpropagation of examples in the \emph{unrolled} grammar (shown in \figref{unrolled}).\footnote{\prose does the unrolling implicitly in its synthesis algorithm.} 
In particular, given examples of an expression $e$ it deduces examples of each argument in $e$ using witness functions.
%\footnote{A witness function for an expression $e$ is essentially a function that encodes the inverse semantics of $e$, i.e., the inverse $e^{-1}$ of $e$.} 
In our case, if $\cellprog$ is chosen to be a $\getcell$ program, the example $\{ (2,2) \mapsto [ (1,2) ] \}$ is translated into the example for the first argument $\cellprog_1$, which is $\{ (2,2) \mapsto [ (1,1) ] \} \disj \{ (2,2) \mapsto [ (1,2) ] \disj \{ (2,2) \mapsto [ (2,2) ] \}$, since we have $\bigcup_{\dir, k}\getcell^{-1}((1,2), \dir, k, \lambda \startcell. \lambda \itercell. \true) = \{ (1,1), (1,2), (2,2) \}$. 
This is shown in \figref{partvsa} as the three edges from the first level $\cellprog$ to the second level $\cellprog_1$, where nodes represent the specifications. 
\prose does backpropagation until it reaches the bottom terminals, i.e., $\inputcell$ in our case, and constructs the atomic version spaces for terminal symbols. 
Then it goes upwards to compose existing version spaces using VSA operations. For instance, node $(2,2) \Rightarrow (2,1)$ for $\cellprog_2$ represents a version space that composes smaller spaces using Union operation. 
As we can see, nodes that represent examples for $\cellprog$, $\cellprog_1$ and $\cellprog_2$ are duplicated, even though they are unrolled from the same symbol $\cellprog$ in the original grammar.

\begin{wrapfigure}{r}{0.35\linewidth}
\vspace{-20pt}
\centering
\includegraphics[scale=0.3]{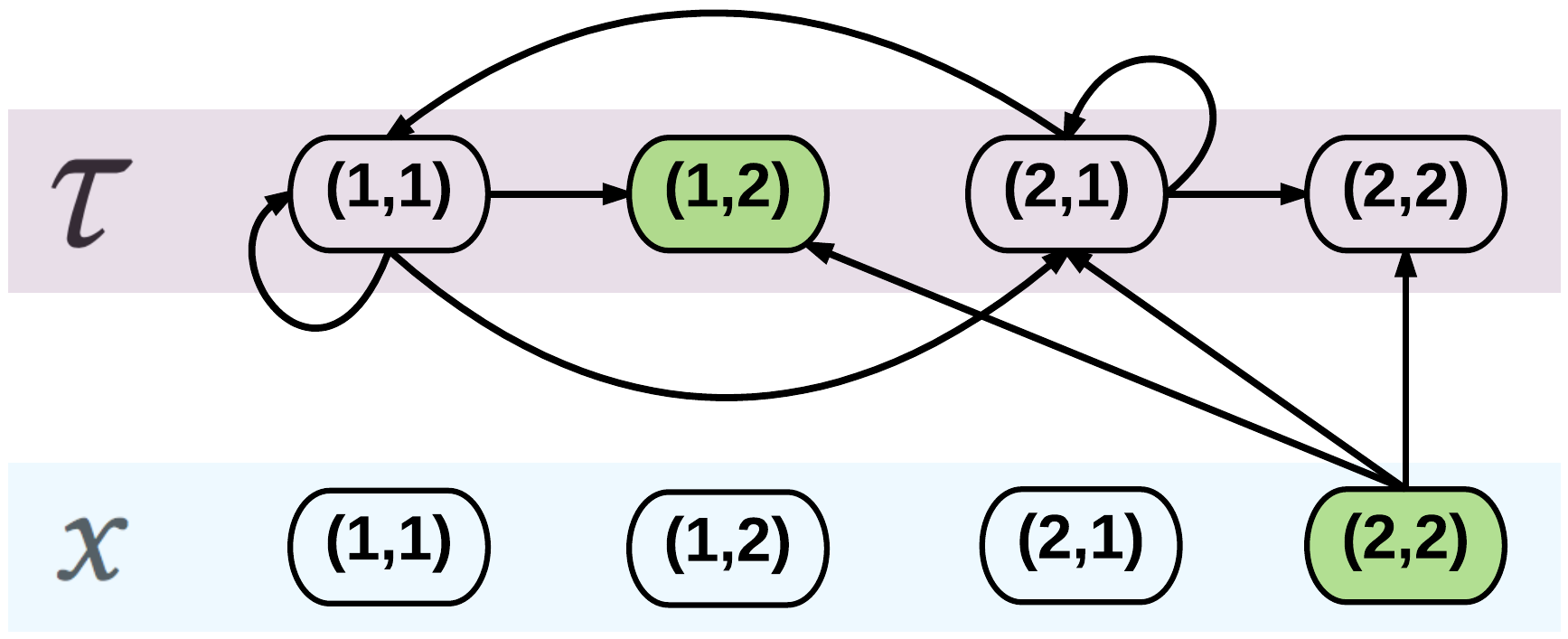}
\caption{Part of the FTA in \tool.}
\figlabel{partfta}
\end{wrapfigure}

In contrast, our FTA technique does not require unrolling, and thus has the potential to create fewer states and lead to a more compact representation. \figref{partfta} shows conceptually how the FTA technique works for the same example. In \figref{partfta}, nodes represent states in the FTA and edges represent transitions. 
Our technique starts from the input example, i.e., $(2,2)$, computes the reachable values using the $\getcell$ construct, and creates transitions from the input value to the output values. It does so until all possible transitions are added. 
As we can see, our technique performs FTA construction in a forward manner, and hence it does not require inverse semantics. Furthermore, it does not require unrolling and results in  a more compact FTA representation in this example.

\end{document}